\documentclass[aps,prb,twocolumn,superscriptaddress,
showpacs,floatfix]{revtex4-1}
\usepackage[caption=false]{subfig}
\usepackage{graphicx}
\usepackage{dcolumn}
\usepackage{bm}
\usepackage{calc} 
\usepackage{amsfonts} 
\usepackage{amssymb} 
\usepackage{amsmath}
\usepackage{subfig} 
\usepackage{natbib}
\usepackage{array,color}
\usepackage{amsbsy}
\usepackage{epstopdf}

\begin{document}

\title{Superconducting and density-wave instabilities of  low dimensional conductors with a Zeeman coupling to a magnetic field}

\author{M. Shahbazi} 
\affiliation{%
Regroupement Qu\'ebecois sur les Mat\'eriaux de Pointe, D\'epartement de physique, Universit\'e de Sherbrooke, Sherbrooke, Qu\'ebec, Canada, J1K-2R1,}%
%\email{maryam.shahbazi@usherbrooke.ca}
\author{Y. Fuseya} 
\affiliation{%
Department of Engineering Science, University of Electro-Communications, Ch Tokyo 182-8585, Japan 
}%
\author{H. Bakrim}
%\email{claude.bourbonnais@usherbrooke.ca }
\affiliation{%
Regroupement Qu\'ebecois sur les Mat\'eriaux de Pointe, D\'epartement de physique, Universit\'e de Sherbrooke, Sherbrooke, Qu\'ebec, Canada, J1K-2R1,}%
\author{A. Sedeki}
%\email{claude.bourbonnais@usherbrooke.ca }
\affiliation{%
Regroupement Qu\'ebecois sur les Mat\'eriaux de Pointe, D\'epartement de physique, Universit\'e de Sherbrooke, Sherbrooke, Qu\'ebec, Canada, J1K-2R1,}%
\author{C. Bourbonnais}
%\email{claude.bourbonnais@usherbrooke.ca }
\affiliation{%
Regroupement Qu\'ebecois sur les Mat\'eriaux de Pointe, D\'epartement de physique, Universit\'e de Sherbrooke, Sherbrooke, Qu\'ebec, Canada, J1K-2R1,}%

\date{\today}

\begin{abstract}
 In the framework of the weak coupling renormalization group technique  we examine the possible instabilities of the extended quasi-one-dimensional electron gas model with both intrachain and interchain electron-electron interactions, including the influence of umklapp scattering and the coupling of spins to a magnetic field. In the limit of purely repulsive intrachain interactions, we confirm the passage from singlet $d$-wave like superconductivity to an inhomogeneous FFLO state under magnetic field. The passage  is accompanied by an anomalous increase  of the upper critical field that scales with the antinesting distance from the quantum critical point joining superconductivity to antiferromagnetism in the phase diagram, as well as the strength of interactions. Adding   weak repulsive interchain interactions promotes the passage from singlet to triplet $f$-wave superconductivity which is expanded under field by the development of a triplet FFLO state with  zero angular momentum projection for the Cooper pairs. The connection between theory and experiments on the anomalous upper critical field in the Bechgaard salts is discussed.
\end{abstract}

\maketitle
\section{Introduction}
The (TMTSF)$_2X$  series of organic conductors, also dubbed the Bechgaard salts series, stands out among the first examples of correlated  electron systems showing the emergence of  superconductivity (SC) following the suppression of a spin-density-wave state (SDW). This is found to occur when either   pressure is applied or by chemical means, from anion  $X$ substitution.\cite{Jerome80,Jerome82,DoironLeyraud09}. This proximity has  fostered  a lot  of debate around the  nature of the SC order parameter in these materials, and in particular its   transformations in   magnetic field  which will   be   the main theoretical focus of the present work.   

The proximity of SC to SDW in the phase diagram  of these quasi-one dimensional  (quasi-1D) materials was soon interpreted as a sign of an intimate connection between both ordered states,  suggesting  that    magnetism is directly  involved  in the development of a SC order parameter. This  led to propose that short-range antiferromagnetic fluctuations of the metallic phase, can act as the source of Cooper  pairing  for electrons \cite{Emery86,BealMonod86,Caron86,Scalapino86,Bourbon88}.  A singlet d-wave  (SCd) gap with nodes on the Fermi surface  was thus predicted  to be the most favourable order parameter for superconductivity, whereas singlet $s$-wave and triplet $p$-wave pairings were found to be  both  suppressed by SDW correlations\cite{BealMonod86}.This was regarded as consistent  with the power law temperature dependence observed in the nuclear spin relaxation rate\cite{Takigawa87,Hasegawa87} and the high sensitivity of superconductivity to impurity scattering\cite{Choi82,Tomic83,Coulon82b,Joo05}. However, the singlet d-wave scenario was later on challenged with the puzzling observation in (TMTSF)$_2$ClO$_4$ of a thermally  activated behaviour of thermal conductivity   below $T_c$ \cite{Belin97}, a behaviour that has  since been found consistent with the penetration depth extracted from muon spin rotation measurements on the same material\cite{Pratt13}.   When combined to  the aforementioned impurity effect \cite{Abrikosov83}, thermal activation may point to a  nodeless  triplet $p$-wave SC gap, clearly in conflict  with the predictions of microscopic calculations. 

On a theoretical basis,  the possibility of triplet SC other than $p$-wave  in purely repulsive quasi-1D electron systems has been examined in different ways. From RPA-like approaches\cite{KurokiGr}, it was found that triplet  $f-$wave  superconductivity (SCf) can compete with SCd if   charge-density-wave (CDW) and SDW fluctuations become of equal importance, a situation that can be reproduced microscopically at sufficiently strong long-range Coulomb interaction along the chains. Such an incursion of SCf besides SCd in the calculated phase diagram  of quasi-1D electron gas model was confirmed by the renormalization group (RG) method  when the long-range part of the Coulomb term dominates other contributions for purely intrachain interactions \cite{Fuseya05b}. When   interchain Coulomb interaction is  included, even weak in amplitude, it was shown from the RG method that bond centered charge-density wave, also called bond-order (BOW) fluctuations  are enhanced besides SDW, which can turn  SCd unstable in favor of a SCf triplet  ordered state \cite{PeryleneSC1}.

In the interval, the triplet scenario for superconductivity in the Bechgaard salts was further promoted from  experiments carried out under magnetic field.  This was borne out by a constant and temperature independent  NMR Knight shift   in the superconducting state of pressurized (TMTSF)$_2$PF$_6$\cite{Lee02}. The violation  from electrical transport measurements of the Glogston criteria or Pauli limit for the critical field of  singlet SC was also understood in terms of triplet pairing  \cite{Brusetti82,Gorkov85,Lee97,Lee95,Lee00}. Resistivity data show the presence of superconductivity up to a critical field $H_{c2}^r$ standing  well above the expected Pauli limiting field $H_P$ known to be bounded by the size of $T_c$ for singlet Cooper pairing.  These experiments were all conducted  for field oriented   in  the $ab'$ plane of highest conduction, an orientation that quenches most of the orbital pair breaking effect, as a result  of the strong anisotropy in the electron motion. In these conditions,   homogeneous superconductivity  can be  sustained at arbitrary field if the SC order parameter  has a triplet character\cite{Lebed86,Dupuis93}. 

 Lower field NMR experiments that were subsequently conducted in (TMTSF)$_2$ClO$_4$   modified this view \cite{Shinagawa07}. They revealed that  the Knight shift  in the superconducting state is actually suppressed in low  field, giving then firm evidence for a singlet SC ground state. However, as the field is increased and crosses some threshold, the Knight shift and  nuclear relaxation rate  recover  their respective normal state values. This  arises while superconductivity persists in electrical transport,  consistently with the aforementioned violation of  the Pauli limit in the $ab'$ plane. 
  
 Theoretically, it was proposed from  various mean-field approaches that NMR and transport experiments could be reconciled if the SC order parameter experiences a singlet to triplet transition  under magnetic field\cite{Shimahara00,Belmechri07,Belmechri08,Aizawa08,Aizawa09}. A transition toward a SCf  state under field  was found to occur using the RG approach to a coupled two-chain version of this problem \cite{Kajiwara09}.  A second  possibility   put forward in the framework of mean-field theory is a transition toward an inhomogeneous FFLO singlet state under field\cite{Lebed86,Dupuis93,Aizawa09,Lebed11}, whose conditions of occurence are particularly optimized for an  open  quasi-1D Fermi surface. The existence of a field-induced FFLO state in the Bechgaard salts has received a certain  empirical support from   the observation of an anisotropic onset of the resistive transition  in  (TMTSF)$_2$ClO$_4$ at $H_{c2}^r$    in the $ab'$ plane\cite{Yonezawa08,Yonezawa08b}. Moreover, recent specific heat experiments performed on  (TMTSF)$_2$ClO$_4$ for similar field orientation \cite{Yonezawa12,Jerome16}, revealed that $H_{c2}^r$ is preceded by a clear thermodynamic signature of the Pauli limit  $H_P$. Besides confirming the  singlet nature of the ground state at low field, this critical field scale corresponds to the transition  seen by NMR under field. 

In a shortened version of the present work, Fuseya {\it et al.,}\cite{Fuseya12}  examined the field dependence of Cooper pairing from the RG approach to  the repulsive quasi-1D electron gas model at incommensurate band filling.  The  magnetic field was exclusively coupled to spins without pair breaking  effects of orbital origin so as to simulate the weakness of the orbital pair breaking for a field oriented in the $ab'$ plane. The calculations revealed that quantum  fluctuations linked to  the  interplay between SDW and SCd  have a sizeable impact on the upper critical field $H_{c2}$.   The $H_{c2}(T)$  critical line  shows a pronounced upturn at low temperature that largely exceeds the predicted Pauli limit.  The difference was found to be  non-universal  for the ratio  $H_{c2}(T)/T_c$, and  SCd was shown to become  unstable against the formation of a d-wave FFLO (dFFLO) state. No indication for field-induced uniform  triplet superconductivity was obtained.

 In the present  work, we carry on the program of Ref.  \cite{Fuseya12}  a step further and extend the RG calculations under magnetic field to the case where half-filling umklapp scattering is present. Umklapp scattering is a key scattering ingredient in systems like the Bechggard salts which presents some half-filling band character. It is also essential in the quantum criticality associated with  the sequence of SDW-SCd instabilities found in these materials\cite{Emery82,Caron86,Bourbon88,Bourbon09,Sedeki12,Shahbazi15,Shahbazi16}. The instability of SCd against the formation of dFFLO state is confirmed under field, together with its strength correlated to the distance to the quantum critical point along the antinesting axis or the strength of interactions. We also investigate the influence of interchain Coulomb interaction in order to examine  if the  singlet to triplet transition induced by this interaction is expanded under field. This is found to be the case with the incursion under field of a triplet $f$-wave FFLO state with zero spin projection for the Cooper pairs, a state that precedes uniform  SCf  type of superconductivity along the interchain interaction axis.
 
In Sec.~II, we introduce the extended quasi-1D electron gas model and the RG method in the presence of a Zeeman coupling of spins to a magnetic field. In Sec.~III, we examine the modification of the phase diagram of the electron gas under magnetic field and the crossover to an inhomogeneous d-wave FFLO in the limit of purely intrachain repulsive interactions. The resulting anomalies in the upper critical field are discussed. In Sec.~IV, the influence interchain repulsive interactions on the possible transitions toward triplet   superconducting orders under field is investigated. We conclude in Sec.~V.
%=======================================================================================
\section{The extended electron gas model  in a magnetic field}
%=======================================================================================
\subsection{Model}
\label{Model}
We consider a linear array of $N_P$ weakly coupled metallic chains of length $L$, separated by the interchain distance $d_b (\equiv 1)$. The partition function is expressed as a functional integral over the anticommuting $\psi's$
\begin{equation}
Z= \int\!\!\int \mathfrak{D}\psi^*\mathfrak{D}\psi\ e^{S_0[\psi^*,\psi] + S_I[\psi^*,\psi]},
\end{equation}
where the quadratic part of the  action is given by
\begin{equation}
S_0[\psi^*,\psi] = \sum_{{\bm{k}},\sigma}\psi^*_{p,\sigma}(\bar{k}) [i\omega_n -E_{p,\sigma}(\bm{k})]\psi_{p,\sigma}(\bar{k}),
\end{equation}
where $\bar{k}=(\bm{k},\omega_n)$, $\bm{k}=(k,k_b)$ is the longitudinal and transverse wave vectors, and $\omega_n$ the fermion Matsubara  frequencies. The spectrum of the electron gas model, in the presence of a Zeeman coupling  of spins  to a magnetic field  $H$,  takes the form
\begin{equation}
\label{Q1DEnergy}
E_{p,\sigma}(\bm{k}) =  v_F (pk-k_F) + \xi_b(k_b) - \sigma h,
\end{equation}
where $p=\pm$ refers to right/left moving carriers along the chains of velocity $v_F$,  with $k_F$  as the 1D Fermi wave vector ($\hbar =1$ and $k_B=1$ throughout).   Here $h = \mu_B H$ and $\sigma =\pm $ is the spin index. 
The transverse part of the electron gas spectrum is
\begin{equation}
 \xi_b(k_b) = -2t_b \cos k_b - 2t_b'\cos 2k_b,
\end{equation}
where $t_b$ is the first nearest-neighbours transverse hopping, whereas the second nearest-neighbour hopping $t'_b\ll t_b$ is the anitinesting tuning parameter that simulates the main effect of pressure in the model.

In the g-ology picture of interactions, the two-body part of the action   can be written in the form
\begin{widetext}
\begin{align}
S_I[\psi^*,\psi]= - {T\over L N_P} \pi v_F \sum_{\{\bar{k},\sigma\}} \Big\{ \  & \ g_\parallel(\boldsymbol{k}_{F,1}^-,\boldsymbol{k}_{F,2}^+;\boldsymbol{k}_{F,3}^-,\boldsymbol{k}_{F,4}^+) \psi^*_{-\sigma}(\bar{k}_1)\psi^*_{+,\sigma}(\bar{k}_2)\psi_{-,\sigma}(\bar{k}_3)\psi_{+,\sigma}(\bar{k}_4) \cr
+ &\, g_{1\perp}(\boldsymbol{k}_{F,1}^-,\boldsymbol{k}_{F,2}^+;\boldsymbol{k}_{F,3}^-,\boldsymbol{k}_{F,4}^+) \psi^*_{-,\sigma}(\bar{k}_1)\psi^*_{+,-\sigma}(\bar{k}_2)\psi_{-,-\sigma}(\bar{k}_3)\psi_{+,\sigma}(\bar{k}_4) \cr\cr
+ &\, g_{2\perp}(\boldsymbol{k}_{F,1}^+,\boldsymbol{k}_{F,2}^-;\boldsymbol{k}_{F,3}^-,\boldsymbol{k}_{F,4}^+) \psi^*_{-,\sigma}(\bar{k}_1)\psi^*_{+,-\sigma}(\bar{k}_2)\psi_{+,-\sigma}(\bar{k}_3)\psi_{-,\sigma}(\bar{k}_4)\cr\cr
+ & \, {1\over 2}\big[\, g_{3\perp}(\boldsymbol{k}_{F,1}^+,\boldsymbol{k}_{F,2}^+;\boldsymbol{k}_{F,3}^{-},\boldsymbol{k}_{F,4}^{-}) \psi^*_{+,\sigma}(\bar{k}_1)\psi^*_{+,-\sigma}(\bar{k}_2)\psi_{-,-\sigma}(\bar{k}_3)\psi_{-,\sigma}(\bar{k}_4) + {\rm c.c.}  \big] \Big\}\cr\cr
& \times \delta_{\bar{k}_1+\bar{k}_2,\bar{k}_3+ \bar{k}_4 (\pm \bar{G})}.
\end{align}
\end{widetext}
The interaction parameters are defined for ingoing and outgoing electrons on the open Fermi surface $\bm{k}_F^p(k_b)=(pk_F(k_b),k_b)$ consisting of two ($p=\pm$) sheets parametrized by $k_b$ from the condition $E_p(\bm{k}^p_F)=0$ in zero field. We have in order, the total backscattering amplitude for parallel spins, $g_\parallel= g_{1\parallel} - g_{2\parallel}$, which incorporates by exchange a forward scattering contribution; the forward scattering for antiparallel spins, $g_{2\perp}$; and  umklapp scattering $g_{3\perp}$ between antiparallel spins  for which the longitudinal lattice vector $\bar{G}=(0,4k_F,0)$ is involved in momentum conservation. All the couplings are dimensionless and normalized by $\pi v_F$.

In the framework of the extended electron gas model\cite{PeryleneSC1,Gorkov74}, the bare interactions  superimpose a purely intrachain contribution and an interchain part between nearest-neighbour chains,
\begin{align}
\label{gialpha}
g_{i,\alpha} (\bar{k}_b)\equiv & \, g_{i,\alpha}(k_{b1},k_{b2},k_{b3}) \cr
= & \,  g_{i,\alpha}+ 2\mathfrak{g}_{i,\alpha}\cos(k_{b1}-k_{b2}),
\end{align}
where $i=1,2,3$ and $\alpha ={\parallel,\perp}$ for the spins orientation. At the bare level, the transverse momentum dependence is coming solely from the interchain coupling, $\mathfrak{g}_{i,\alpha}$, a dependence that is modified on the Fermi surface  by the RG flow of the coupling constants.  

We will fix the range of the main parameters of the above model in order to simulate the experimental phase diagram of  the Bechgaard salts in zero field.  From band calculations\cite{Grant83,Ducasse86}, we shall take $E_F= v_Fk_F= 3000$K for typical range of  longitudinal Fermi energy and $t_b=200$K for the amplitude of the transverse hopping along the $b$ direction. The antinesting amplitude  $t_b'$ of the spectrum will be kept small compared to $t_b$ and will serve as a tuning parameter to mimic the effect of pressure. As for interactions, although it exists a large range of possible values able to generate a zero field phase diagram compatible with observations for the Bechgaard salts, we can follow  the  arguments of earlier works to obtain a reasonable set of figures for the intrachain couplings \cite{PeryleneSC1,Bourbon09,Sedeki12,Shahbazi15,Shahbazi16}. For instance, the bare intrachain  backscattering amplitude can be fixed to
 $ g_{1,\alpha}\simeq 0.32$, consistently with the range of values extracted from the enhancement of  uniform susceptibility measurements\cite{Wzietek93}. The presence of a small dimerization gap {$\Delta_D\ll E_F$}, in the middle of an otherwise three-quarter filled band\cite{Grant83,Ducasse86}, leads to  weak half-filling umklapp scattering, $g_{3\perp} \approx g_{1\perp} \Delta_D/E_F$\cite{Barisic81,Emery82,Penc94}. This gives for umklapp the range of values $g_{3\perp}\approx 0.02...0.03$. The bare forward scattering amplitude can  then be adjusted to  $ g_{2,\alpha}\simeq 0.64$, so  that the calculated temperature scale  of the SDW instability from RG at relatively low antinesting falls in the range of observed values $T_{\rm SDW} \sim 10$K for the Bechgaard salts at low pressure\cite{Jerome82}. With the above  figures,  a   SDW to SCd  sequence of instabilities can be  obtained by the RG (e.g., $h=0$ critical line of Fig.~\ref{Phases_intra} obtained, at $g_{3\perp} =0.025$), which is compatible with experiments\cite{Jerome82,DoironLeyraud09}.     Finally, regarding the amplitudes of   repulsive interchain interaction $\mathfrak{g}_{i,\alpha}$,    they will be taken variable,  but kept small   in comparison to their respective intrachain counterparts $g_{i,\alpha}$.

\subsection{Renormalization group equations}

We apply a Kadanoff-Wilson RG approach to the  extended quasi-1D  electron model introduced in  the previous subsection. The approach, which has been detailed in previous works\cite{PeryleneSC1,Sedeki12,Fuseya12} consists in the perturbative successive partial integrations of electron degrees of freedom in the partition function $Z$ on  energy shells of thickness $\Lambda(\ell)d\ell$ at energy distance $\Lambda(\ell) =\Lambda_0 e^{-\ell}$ above and below the Fermi surface, where $\Lambda_0 \equiv E_F$ is the initial cutoff fixed at the Fermi energy.  Each energy shell is divided into $N_p$ patches, in which a transverse momentum integration is carried out for the internal  variables of the logarithmically singular electron-electron (Copper) and electron-hole (Peierls) loops of the scattering channels.

At the one-loop level, the RG flow equations for the 3-momentum dependent scattering amplitudes $g_{i,\alpha}$ at non zero magnetic field are
\begin{widetext}
\begin{align}\label{gflow}
\partial_\ell g_\parallel(\bar{k}_b) =  & - \langle\, g_\parallel(\bar{k}_{b1}) g_\parallel(\bar{k}_{b2}) \mathcal{I}_{P}^0(k_b , q_{P})\,\rangle_{k_b}
- \langle\,g_{1\perp}(\bar{k}_{b1}) g_{1\perp}(\bar{k}_{b2}) \mathcal{I}_{P}^{4h}(k_b , q_{ P})\,\rangle_{k_b} \cr
&+  \langle\, g_\parallel(\bar{k}_{b3}) g_\parallel(\bar{k}_{b2}) \mathcal{I}_{C}^0(k_b , q_{C})\,\rangle_{k_b}
-   \langle\, g_{3\perp}(\bar{k}_{b1}) g_{3\perp}(\bar{k}_{b2}) \mathcal{I}_{P}^{4h}(k_b , -q_{P})\,\rangle_{k_b}\cr\cr
\partial_\ell g_{1\perp}(\bar{k}_b) = &-   \langle\, \big[g_\parallel(\bar{k}_{b1}) g_{1b}(\bar{k}_{b2}) 
+  g_\parallel(\bar{k}_{b2}) g_{1\perp}(\bar{k}_{b3})\big] \big(\mathcal{I}_{P}^{0}(k_b , -q_{b P}) + \mathcal{I}_{P}^{4h}(k_b , - q_{P})\big)/2 \,\rangle_{k_b}\cr
&+ \langle\, \big[g_{2\perp}(\bar{k}_{b3}) g_{1\perp}(\bar{k}_{b1}) 
+g_{2\perp}(\bar{k}_{b4}) g_{1\perp}(\beta_{b1})\big] \big(\mathcal{I}_{C}^{4h}(k_b , q_{b C}) + \mathcal{I}_{C}^{0}(k_b , q_{C})\big)/2\,\rangle_{k_b} ,\cr\cr
\partial_\ell g_{2\perp}(\bar{k}_b) =&  -   \langle\,g_{1\perp}(\bar{k}_{b3}) g_{1\perp}(\bar{k}_{b4}) \mathcal{I}_{C}^{4h}(k_b,q_{C})\,\rangle_{k_b}
+ \langle\,g_{2\perp}(\bar{k}_{b3}) g_{2\perp}(\bar{k}_{b4}) \mathcal{I}_{C}^{0}(k_b,q_{C})\,\rangle_{k_b} \cr
&-\langle\, \big[g_{2\perp}(\bar{k}_{b1}) g_{2\perp}(\bar{k}_{b3}) 
+ g_{3\perp}(\bar{k}_{b1}) g_{3\perp}(\bar{k}_{b2})\big] \mathcal{I}_{P}^{0}(k_b , - q_{P}) \,\rangle_{k_b} ,\cr\cr
\partial_\ell g_{3\perp}(\bar{k}_b) = &-\langle \, g_\parallel(\bar{k}_{b1}) g_{3\perp}(\bar{k}_{b2}) \big(\mathcal{I}_{P}^0(k_b , q_{P}) + \mathcal{I}_{P}^{4h}(k_b , q_{P})\big)/2 \,\rangle_{k_b}\cr
&+\langle\, g_\parallel(\bar{k}_{b2}) g_{3\perp}(\bar{k}_{b1}) \big(\mathcal{I}_{P}^{4h}(k_b , -q_{P}) + \mathcal{I}_{P}^0(k_b , -q_{P})\big)/2 \,\rangle_{k_b}\cr
&-\langle \, g_{2\perp}(\bar{k}_{b5}) g_{3\perp}(\bar{k}_{b6}) \mathcal{I}_{P}^0(k_b , q_{P}^\prime) 
+ g_{2\perp}(\bar{k}_{b5}) g_{3\perp}(\bar{k}_{b6})\mathcal{I}_{P}^0(k_b , -q_{P}^\prime) \,\rangle_{k_b},
\end{align}
\end{widetext}
where $\langle \ldots\rangle_{k_b} = 1/N_P\sum_{k_b} \ldots$ and 
\begin{align*}
\bar{k}_{b1} = (k_b , k_{b4} , k_{b1})\\
\bar{k}_{b2} = (k_b , k_{b2} , k_{b3})\\
\bar{k}_{b3} = (k_{b1} , k_{b2} , k_b)\\
\bar{k}_{b4} = (k_{b3} , k_{b4} , k_b)\\
\bar{k}_{b5} = (k_b , k_{b4} , k_{b2})\\
\bar{k}_{b6} = (k_{b1} , k_b , k_{b3})\\
\end{align*}
$q_{P}^{(\prime)}= k_{b3}-k_{b2,1}=k_{b1,2}-k_{b4}$ and 
$q_{C}= k_{b1,3}+k_{b2,4}$.
The on-shell Peierls ($\nu=P$) and Cooper ($\nu=C$) loops at finite $T$ and $h$ are given by 
\begin{align}\label{Afunction}
 {\cal I}_{\nu}^{\kappa h}&(k_b,q_{\nu}^{(\prime)}) =  \frac{\Lambda(\ell)}{2 } \sum \limits_{\lambda=\pm 1} \int_{k_b - \frac{\pi}{N_P}}^{k_b + \frac{\pi}{N_P}} {dk_b\over 2\pi} \cr
& \times \dfrac{\theta(\vert \Lambda(\ell) + \lambda A_\nu^{\kappa h}\vert - \Lambda(\ell))}{2 \Lambda(\ell)+\lambda A_\nu^{\mu h}} \cr
&  \times   \Big[ \tanh[\beta \Lambda(\ell)/2] + \tanh[\beta( \Lambda(\ell)/2 + \lambda A_\nu^{\kappa h} /2)]\Big],\cr
\end{align}
where for the loop field dependence, $\kappa=0,4$. Here $\theta(x)$ is the Heaviside function $[\theta(0) \equiv {1\over 2}]$, and
\begin{align}
A_\nu^{\kappa h}(k_b,q_{\nu}^{(\prime)})&=- \xi_b(k_b) - \eta_\nu \xi_b(\eta_\nu k_b+q_{\nu}^{(\prime)})\cr
& + \eta_\nu \xi_b(\eta_\nu k_{b2(4)}+q_{\nu}^{(\prime)}) + \xi_b(k_{b2(4)}) + \kappa h, \cr
\end{align}
for which $\eta_{P,C}=\pm 1$.

To find out the nature of instabilities of the electron system, we compute the susceptibilities associated with the different  possibilities of staggered density-wave and  Cooper pairing correlations. Under successive partial integrations of the RG transformation, the linear coupling of pair of carriers to an external source field  $h_\mu$ in the correlation channel $\mu$, yields the generic expression of the normalized temperature dependent susceptibility ($\tilde{\chi}_\mu=\pi v_F \chi_\mu$) at the wave vector $\bm{q}_\mu$:
\begin{equation}
\label{Kimu}
\tilde{\chi}_\mu(\bm{q}_\mu) = 2\int_0^\infty \langle z^2_\mu(k_b) {\cal I}^{\kappa_\mu h}_\mu(k_b,q_\mu)\rangle_{k_b}d\ell,
\end{equation}
where $z_\mu(k_b)$ is the renormalization factor for the source-pair vertex. It obeys the flow equation
\begin{equation}
\partial_\ell z_\mu(k_b) = {1\over 2}\langle f_\mu(k'_b)g_\mu(\bar{k}'_b) {\cal I}^{\kappa_\mu h}_\mu(k'_b,q_\mu)\rangle_{k'_b}.
\end{equation}
where $g_\mu$ is a momentum dependent combination of couplings for the  correlation of the channel $\mu$ and $f_\mu(k_b)$ is a form factor associated with the nature of correlations.  

If we first look   at the density-wave   susceptibilities for which $f_\mu=1$, we  have in the charge sector,   the site-centred ({$\mu=$ CDW}) and bond-centred  ({$\mu=$ BOW}) charge-density wave  susceptibilities, corresponding to the following combinations of couplings at the modulation (nesting) wave vector  $\bm{q}_{\rm CDW}=\bm{q}_{\rm BOW}= (2k_F,\pi)$,
\begin{align}
\label{CDWBOW}
& g_\mu {\cal I}^{\kappa_\mu h}_\mu  \big|_{\mu ={\rm CDW,BOW}}=   -\big [g_{1\perp}(k^\prime_b  + \pi , k_b , k_b + \pi) \ + \cr
   &   g_\parallel(k_b^\prime + \pi , k_b , k_b + \pi) 
\pm g_{3\perp}(k_b^\prime + \pi, k_b + \pi , k_b)\big] \mathcal{I}_{P}^{2h}(k'_b,\pi). 
\end{align}
In the spin sector, the site-centred SDW susceptibility is likely to become singular. In presence of a magnetic field along  $z$, the rotational symmetry is broken, which splits this susceptibility into  longitudinal ($\tilde{\chi}_{{\rm SDW}_z}$) and transverse ($\tilde{\chi}_{{\rm SDW}_{xy}}$) components  for which, 
 \begin{align}
\label{SDWmu}
& g_\mu {\cal I}^{\kappa_\mu h}_\mu  \big|_{\mu ={\rm SDW}_z,{\rm SDW}_{xy}}=    \big[g_{2\perp}(k^\prime_b + \pi , k_b, k_b + \pi ) \cr 
&+  g_{3\perp}(k^\prime_b + \pi, k_b + \pi, k^\prime_b)\big]  \mathcal{I}_{P}^{2h,0}(k'_b,\pi)
\end{align}

If we consider in the second place the SC susceptibilities at $\bm{q}_\mu=0$ that may be potentially singular in the presence of a magnetic field, 
we have for the singlet SC channel,
   \begin{align}
\label{SCsinglet}
& f_\mu g_\mu {\cal I}^{\kappa_\mu h}_\mu =    - f_\mu(k_b')\big[g_{1\perp}(k'_b,-k_b',k_b) \cr 
&+  g_{2\perp}(k'_b,-k_b',k_b)\big]  \mathcal{I}_{C}^{2h}(k'_b,0).
\end{align}
For singlet s-wave susceptibility,   $\tilde{\chi}_{{\rm SS}}$, $f_{\rm SS} =1$; for d-wave susceptibility, $\tilde{\chi}_{{\rm SCd}}$,  $f_{\rm SCd}(k_b) = \sqrt{2} \cos k_b$; for $g$-wave,  $\tilde{\chi}_{{\rm SCg}}$,  $f_{\rm SCg}= \sqrt{2}\cos 2k_b$, etc.

For the triplet channel at $\bm{q}_\mu=0$, the SC susceptibilities are governed by the expressions 
   \begin{align}
\label{SCtriplet0}
& f_\mu g_\mu {\cal I}^{\kappa_\mu h}_\mu =     f_\mu(k_b')\big[g_{1\perp}(k'_b,-k_b',k_b) \cr 
&-  g_{2\perp}(k'_b,-k_b',k_b)\big]  \mathcal{I}_{C}^{2h}(k'_b,0),
\end{align}
for antiparallel spins at $S_z=0$, whereas for parallel spins at $S_z=\pm 1$,
   \begin{align}
\label{SCtriplet1}
  f_\mu g_\mu {\cal I}^{\kappa_\mu h}_\mu = f_\mu(k_b') g_{\|}(k'_b,-k_b',k_b) \mathcal{I}_{C}^{0}(k'_b,0),
\end{align}
For both cases, we have  for $p$-wave susceptibility, $\tilde{\chi}_{{\rm SCp}}$, $f_{\rm SCp}=1$; $f$-wave $\tilde{\chi}_{{\rm SCf}}$,  $f_{\rm SCf}= \sqrt{2}\cos k_b$; etc.

Now for superconductivity, it is possible for electrons of opposite spins to form Cooper pairs  with a nonzero  momentum $\bm{q}_h = (2h/v_F, 0)$ in a FFLO state.  This case requires a separate treatment of the pair vertex  $z_\mu$\cite{Fuseya12}, which actually   splits into $z^{\uparrow\downarrow}_\mu$ and $z^{\downarrow\uparrow}_\mu$ for opposite spins. For singlet FFLO, these are governed by
\begin{align}\label{z_ssq}
&\partial_\ell z_{\mu}^{\uparrow\downarrow(\downarrow\uparrow)}(k_b) = -\Big <f_\mu(k_b') \big[g_{1\perp}(-k^\prime_b , k^\prime_b,- k_b) z_{\mu}^{\downarrow\uparrow(\uparrow\downarrow)}(k^\prime_b) \cr
&\times \mathcal{I}_{C}^{0(4h)}(k'_b,0)
+ g_{2\perp}(-k^\prime_b , k^\prime_b, - k_b)z_{\mu}^{\uparrow\downarrow(\downarrow\uparrow)}(k^\prime_b)\cr
&\times  \mathcal{I}_{C}^{4h(0)}(k'_b,0)\big] \Big >_{k^\prime_b}, 
\end{align}
where for $s$-wave FFLO, $f_{\rm sFFLO}=1$; d-wave FFLO, $ f_{\rm dFFLO}(k_b)=\sqrt{2}\cos k_b$; etc.

 It is also possible in principle for triplet Cooper pairing with zero total spin projection, $S_z=0$,  to develop an inhomogeneous FFLO state following the equations
\begin{align}\label{z_tsq}
&\partial_\ell z_{\mu}^{\uparrow\downarrow(\downarrow\uparrow)}(k_b) = \Big <f_\mu(k_b') \big[g_{1\perp}(-k^\prime_b , k^\prime_b,- k_b) z_{\mu}^{\downarrow\uparrow(\uparrow\downarrow)}(k^\prime_b) \cr
&\times \mathcal{I}_{C}^{0(4h)}(k'_b,0) 
- g_{2\perp}(-k^\prime_b , k^\prime_b, - k_b)z_{\mu}^{\uparrow\downarrow(\downarrow\uparrow)}(k^\prime_b) \cr
& \times \mathcal{I}_{C}^{4h(0)}(k'_b,0)\big] \Big >_{k^\prime_b}, 
\end{align}
where for \hbox{$S_z=0$} of $p$-wave, \hbox{$f_{\rm pFFLO}=1$}; $f$-wave, \hbox{$f_{\rm fFFLO}=\sqrt{2}\cos k_b$}; etc.

The corresponding  temperature dependent  susceptibilities for  the whole set of FFLO states take  the following form  
\begin{align}\label{ki_ssq}
\tilde{\chi}_{\mu}(\bm{q}_h) = \int_0^\infty\langle \big[&{\ [z_{\mu}^{\uparrow \downarrow}(k_b)]}^2 \mathcal{I}_{C \ell}^{0}(k_b,0) \cr+ &\ {[z_{\mu}^{\downarrow \uparrow}(k_b)]}^2 \mathcal{I}_{C \ell}^{4h}(k_b,0) \big]\rangle_{k_b} d\ell.
\end{align}

%=======================================================================================
\section{Results for the model with intrachain interactions}
\begin{figure} 
 \includegraphics[width=9cm]{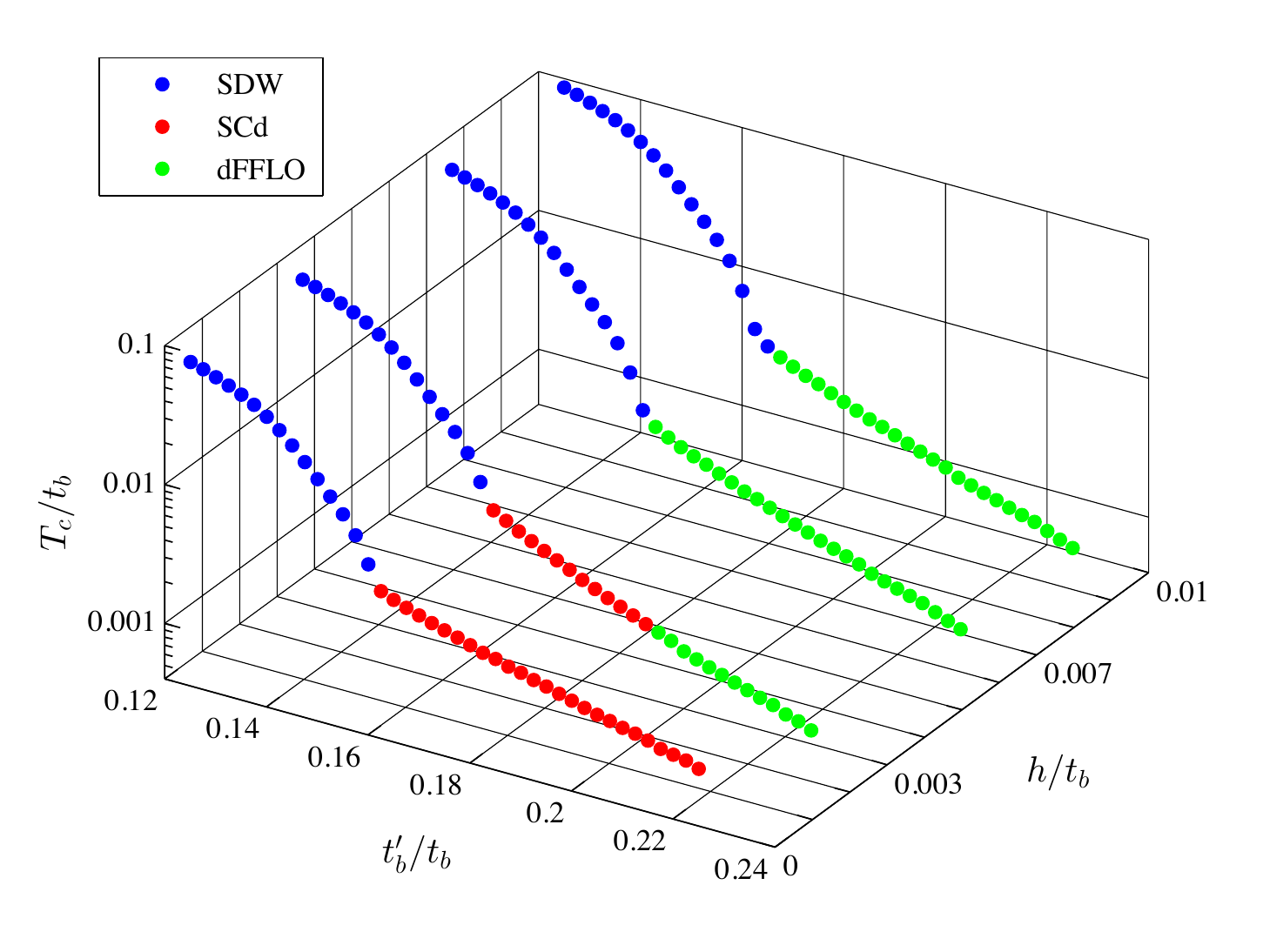}
 \caption{ The phase diagram of the  quasi-1D electron gas model with intrachain interactions   as a function  of magnetic field. \label{Phases_intra} }
 \end{figure}

\begin{figure}  
\includegraphics[width=8cm]{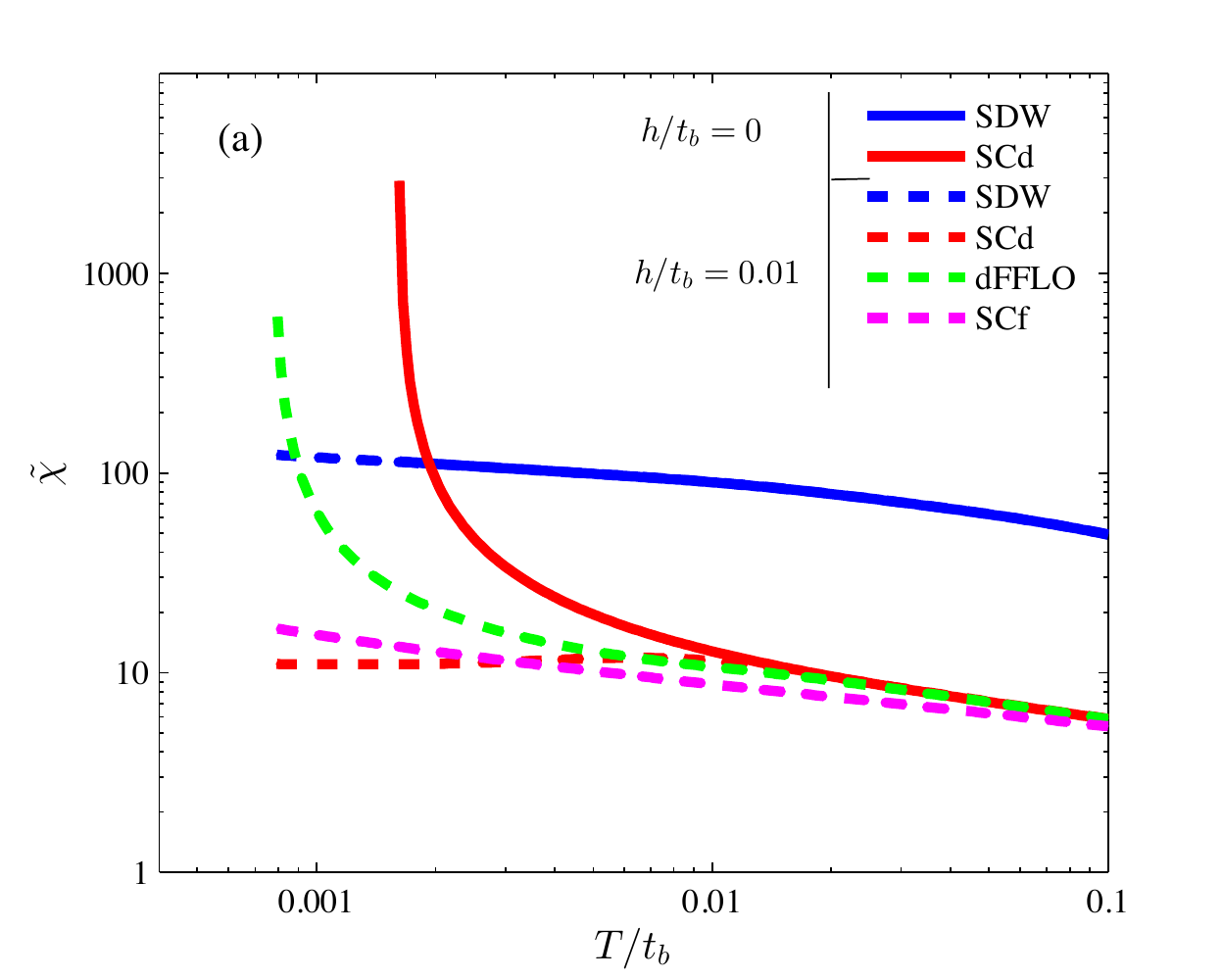} \\
\includegraphics[width=8cm]{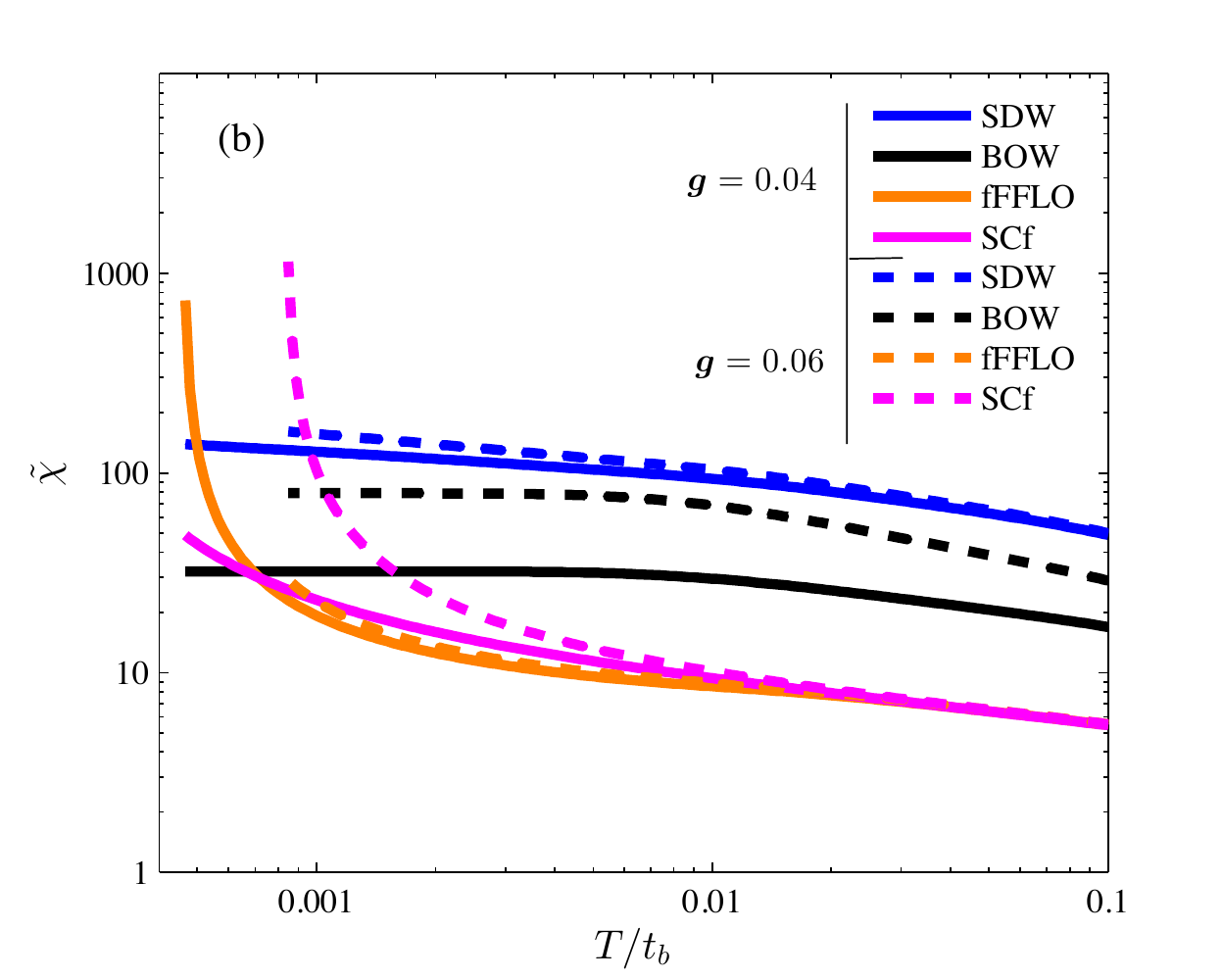}
\caption{ Temperature dependence of the susceptibilities in the normal phase of  the superconducting sector of the phase diagram at  $t^\prime_b / t_b = 0.21$. (a): For intrachain couplings only, $ \mathfrak{g}_{i}=0$ at zero and finite magnetic fields. (b): For finite interchain couplings $ \mathfrak{g}_{i}\ne0$  at  $h/t_b= 0.01$.}
\label{Suscg0}
\end{figure}

\begin{figure}
     \includegraphics[width=5.5cm]{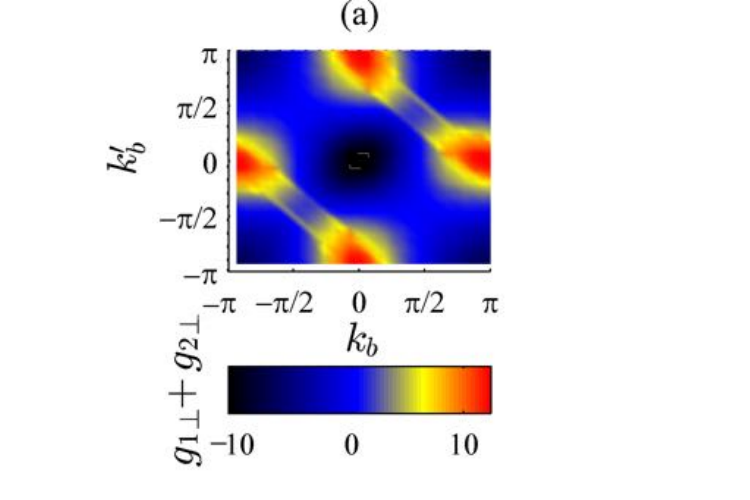}\\ 
     \includegraphics[width=5.5cm]{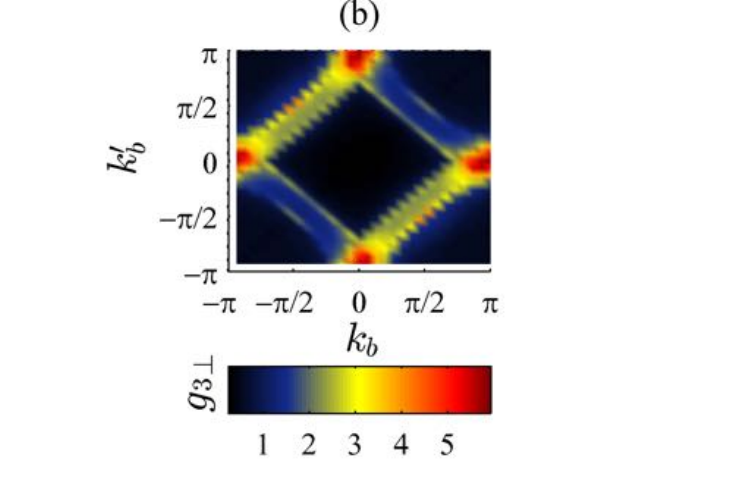}\\
     \includegraphics[width=5.5cm]{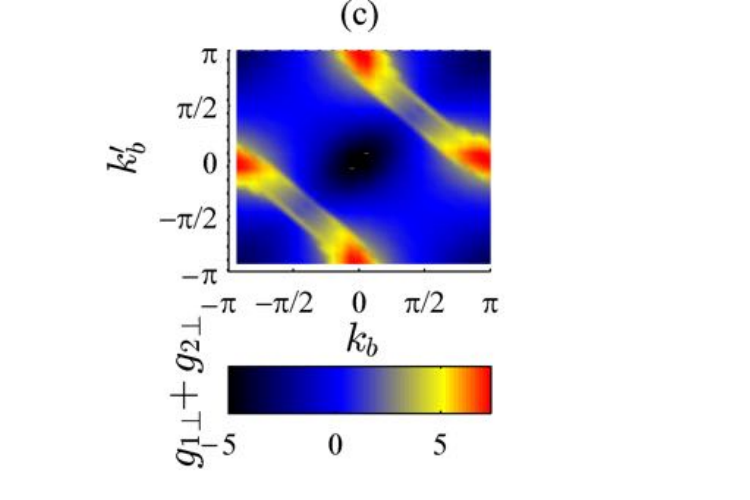} 
  \caption{Low temperature renormalized effective scattering amplitudes at  $t^\prime_b / t_b = 0.21$ for singlet Cooper pairing $g_{1\perp}(k_{b}', -k_{b}', k_{b}) +g_{2\perp}(k_{b}', -k_{b}', k_{b})$ in the $k_{b},k_{b}'$ plane for the normal phase of (a) :  SCd ($h = 0$); (b): umklapp amplitude  $g_{3\perp}(k_{b}', -k_{b}', k_{b})$ at $h=0$; (c) : dFFLO ($h / t_b = 0.01$). 
  \label{gintra}}
\end{figure}

%=======================================================================================
We first examine the results of integration of the above  RG equations  for $\mathfrak{g}_{i,\alpha}=0$ in Eq.~(\ref{gialpha}), namely when  only intrachain interactions are present. In  zero magnetic  field the sequence of instabilities obtained for the input parameters of model given in Sec.~\ref{Model} coincides with the one found in previous works\cite{Shahbazi15,Shahbazi16}. Thus, at relatively low antinesting amplitude $t_b'$ , a singularity in $\tilde{\chi}_{\rm SDW}$ at $\bm{q}_{\rm SDW}=(2k_F,\pi)$  is found  from (\ref{Kimu}) and (\ref{SDWmu}). The critical temperature $T_{\rm SDW}$ is traced   in Fig.~\ref{Phases_intra},  which decreases monotonically by increasing $t_b'$. Close to the critical value $t_b'^*$ ($\simeq 32$K, $t_b'^*/t_b\simeq 0.16$), $T_{\rm SDW}$ drops rapidly until  $t_b'^*$ is reached and  the system  becomes unstable against  the formation of a SCd state, with the divergence of $\tilde{\chi}_{\rm SCd}$ coming from the singularity of (\ref{SCsinglet}) at a maximum $T_c\sim 1$K [See Fig.~\ref{Suscg0}~(a)]; $T_c$ then closes the sequence by its steady decrease with $t_b'>  t_b'^*$, as shown in  Fig.~\ref{Phases_intra}. The typical  momentum profile of the SC combination of couplings $g_{\rm SCd}(k_b',-k_b',k_b)$  in the $k_bk_b'$ plane for temperature close to $T_c$, plane shows  pronounced modulations compatible with the form factor $f_{\rm SCd}$ for SCd superconductivity. According to Fig.~\ref{gintra}~(b), this  modulation in momentum space is intimately connected   with the amplitude and anisotropy developed by  umklapp, $g_{3\perp}(k_b',-k_b',k_b)$, along the lines $k_b'= \pm k_b \pm \pi$, and which from (\ref{SDWmu}), is directly involved in  the strength of SDW correlations  responsible for SCd pairing.

%~~~~~~~~~~~~~~~~~~~~~~~~~~~~~~~~~~~~~~~~~~~~~~~~~~~~~~~~~~~~~~~~~~~~~~~~~~~~~~~~~~~~~~~
 %~~~~~~~~~~~~~~~~~~~~~~~~~~~~~~~~~~~~~~~~~~~~~~~~~~~~~~~~~~~~~~~~~~~~~~~~~~~~~~~~~~~~~~

For non zero magnetic field, the SDW instability at low $t_b'$ is now taken place for spins oriented in the $xy$ plane transverse to the field. The amplitude of $T_{\rm SDW}$ obtained in  Fig.~\ref{Phases_intra} is slightly reinforced compared to  zero field. This reinforcement of antiferromagnetism   agrees with an increase of the critical  $t_b'^*$ with $h$. At very low field, this presents as an increase of the maximum SCd $T_c$ with $h$, which is possible because orbital pair breaking is absent from the model. 
However, the  SDW$\to$SCd sequence of instabilities is rapidly altered under field.  As shown in Fig.~\ref{Phases_intra}, where an incursion of a  dFFLO instability takes place along the antinesting axis, as signalled  by a  singularity of $\tilde{\chi}_{\rm dFFLO}(\bm{q}_h) $ coming from (\ref{z_ssq}) at a nonzero pairing momentum $\bm{q}_h = (2h/v_F, 0)$. The related divergence occurs at a $T_c$ that is steadily suppressed under field, but whose amplitude is significantly enhanced compared to   mean-field calculations in which the interplay between the Cooper and density-wave pairing is neglected\cite{Fuseya12,Aizawa09}. From Fig.~\ref{gintra}~(c), the combination of couplings  $g_{1\perp}(k_b',-k_b',k_b) + g_{2\perp}(k_b',-k_b',k_b)$ entering in (\ref{z_ssq}) for singlet FFLO superconductivity presents also d-wave like modulations in the   $k_bk_b'$ plane, but of weaker amplitude compared to the zero field situation.

Regarding triplet superconductivity, we see from Fig.~\ref{Suscg0} (a) that apart  from a regular  enhancement of  $\chi_{\rm SCf}$ at low temperature no crossover to triplet superconductivity is found under field when only intrachain interactions are present. This result differs from the mean-field phenomenology when both singlet and triplet pairing interactions are present \cite{Shimahara00,Belmechri07,Belmechri08}

%~~~~~~~~~~~~~~~~~~~~~~~~~~~~~~~~~~~~~~~~~~~~~~~~~~~~~~~~~~~~~~~~~~~~~~~~~~~~~~~~~~~~~~ 
%~~~~~~~~~~~~~~~~~~~~~~~~~~~~~~~~~~~~~~~~~~~~~~~~~~~~~~~~~~~~~~~~~~~~~~~~~~~~~~~~~~~~~~

%~~~~~~~~~~~~~~~~~~~~~~~~~~~~~~~~~~~~~~~~~~~~~~~~~~~~~~~~~~~~~~~~~~~~~~~~~~~~~~~~~~~~~~~

\subsection{$H$-$T$ phase diagram}
\begin{figure} 
 \includegraphics[width=8cm]{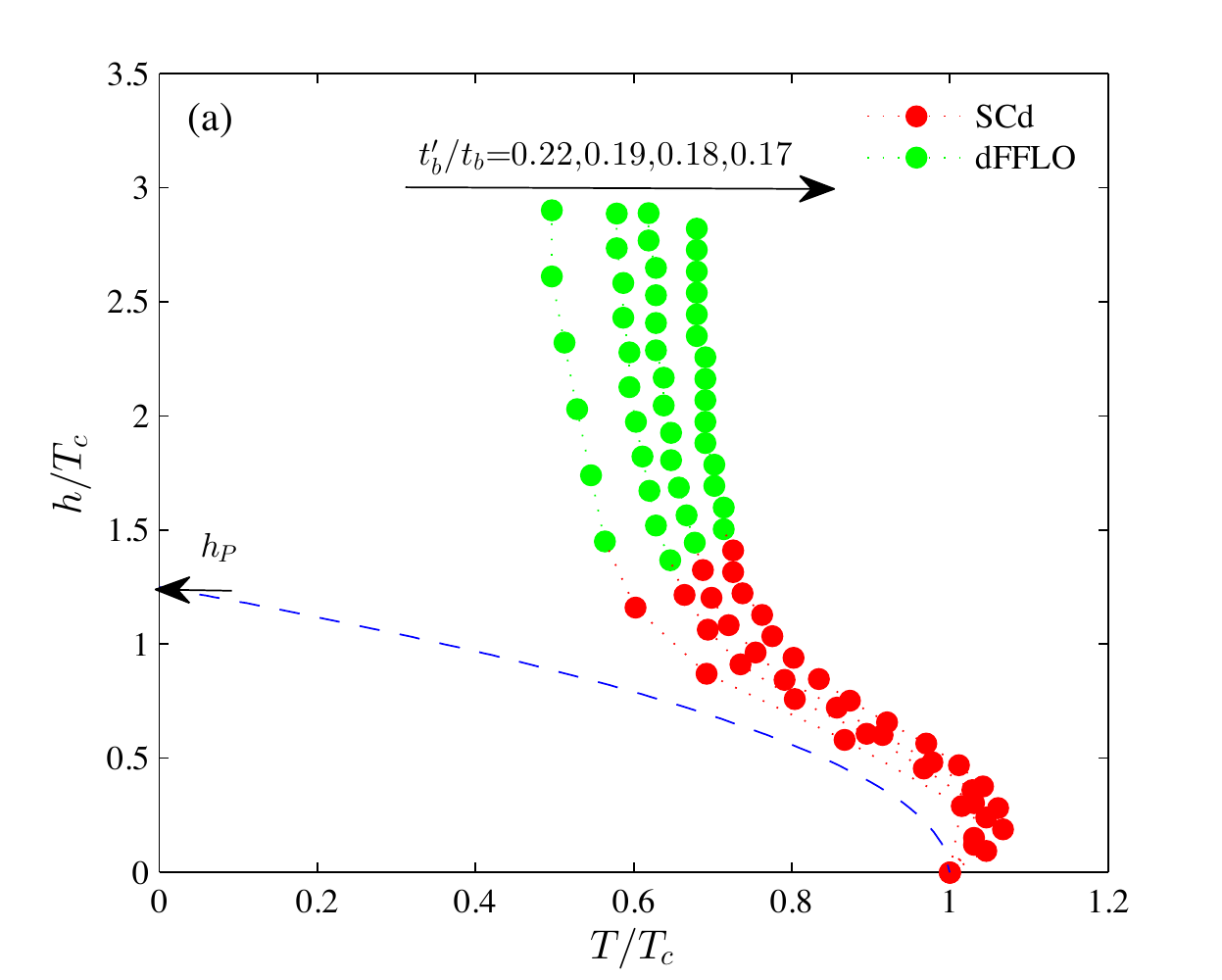}\\
  \includegraphics[width=8cm]{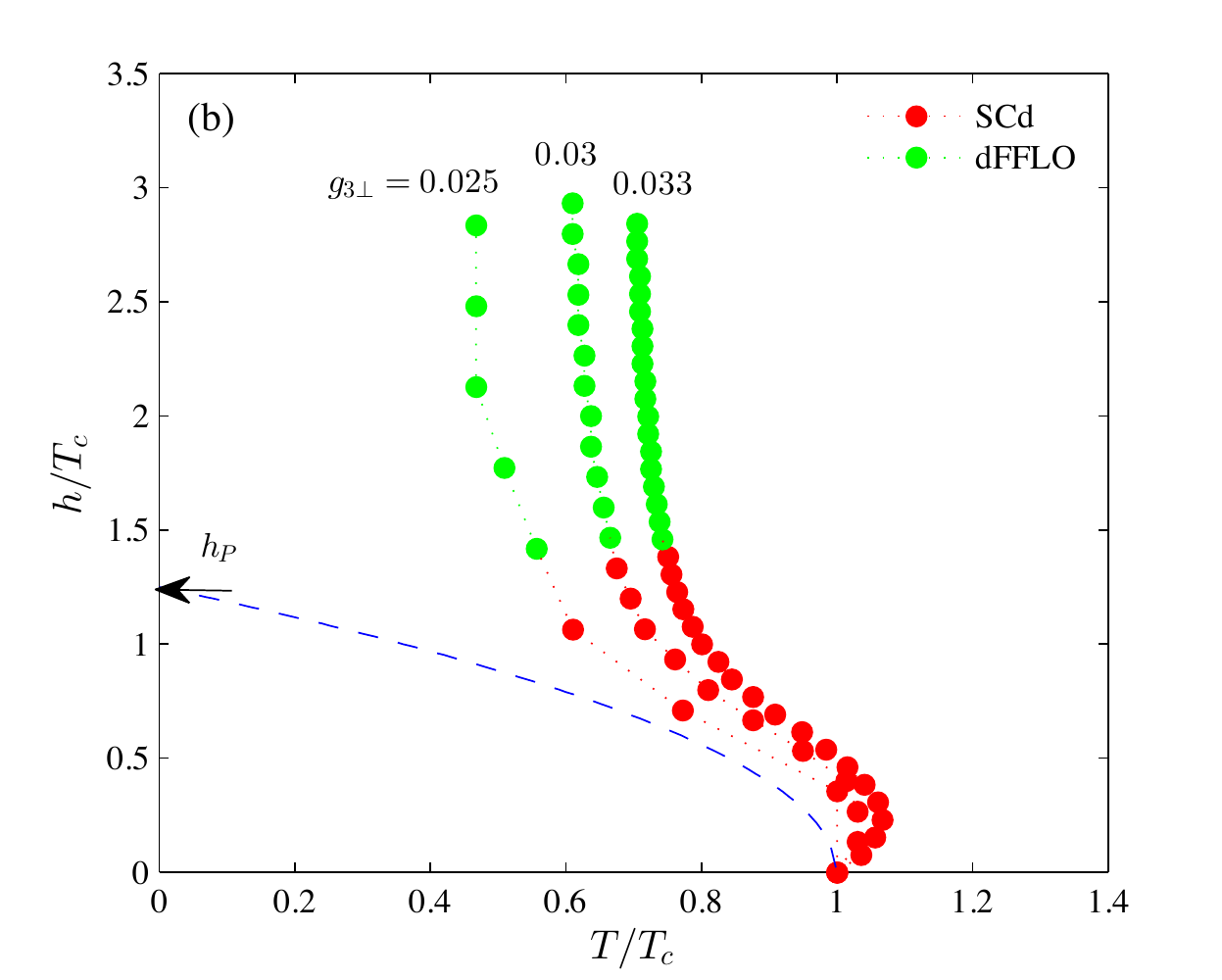}\\
 \caption{ The variation of the  normalized upper critical field  $h_{c2}/T_c$ with temperature for (a) : various $t^\prime_b$ and (b) : different amplitudes of umklapp scattering $g_{3\perp}$ at $t_b'/t_b=0.22$. The dashed line is mean-field result for the upper critical field in the Pauli limit. \label{hc2_intra} }
 \end{figure}

In the superconducting sector of the phase diagram of Fig.~\ref{Phases_intra}, one can follow the critical temperature $T_c(h)$ with field, or conversely the upper critical field profile $h_{c2}(T)$ with temperature of  Fig.~\ref{hc2_intra}. At very low field, the slope $d h_{c2}/dT$ is at first positive, indicating that $T_c$ increases with $h$.  This results from the strengthening of SDW correlations, which  as  the source of Cooper pairing in the SCd channel, exceeds the  field  pair breaking effect on the singlet state  in (\ref{Kimu}) and (\ref{SCsinglet}). As previously mentioned, this enhancement of $T_c$ takes place because  orbital  effect caused by the  field is absent in the present model. 
 %~~~~~~~~~~~~~~~~~~~~~~~~~~~~~~~~~~~~~~~~~~~~~~~~~~~~~~~~~~~~~~~~~~~~~~~~~~~~~~~~~~~~~~
%~~~~~~~~~~~~~~~~~~~~~~~~~~~~~~~~~~~~~~~~~~~~~~~~~~~~~~~~~~~~~~~~~~~~~~~~~~~~~~~~~~~~~~~  
  At higher field, however, singlet pair breaking dominates and $T_c$ decreases, as shown in Fig.~\ref{hc2_intra}. The values of $h_{c2}$  are systematically above the mean-field result for the pure Pauli limit (dashed lines of  Fig.~\ref{hc2_intra})\cite{Matsuda07}.
   Instead of extrapolating to a field close to $h_P$ in the low temperature limit, $h_{c2}$ continues until a crossover to  an inhomogeneous $d$FFLO state is achieved. In the dFFLO regime, $h_{c2}$ not only  exceeds  the Pauli limiting field $h_P (\simeq 1.25 T_c$), but also the Pauli  limit  of FFLO state  for isotropic 2D ($h_P \simeq 1.78  T_c$) and 3D ($h_P \simeq 1.34 T_c$) superconductors \cite{Matsuda07}. 
   
   Another important feature of the present results\cite{Fuseya12}, which  contrasts with mean-field type of calculations, is the non-universality of  the ratio $h_{c2}/T_c$, as a function of either the  antinesting amplitude $t_b'$  [Fig.~\ref{hc2_intra}~(a)] or interaction [e.g., $g_{3\perp}$, in Fig.~\ref{hc2_intra}~(b)]. At the root of this lack of universality stands  SDW fluctuations  as the source of Cooper pairing. In this respect, the RG flow equations (\ref{gflow}) tell us that, in contrast to mean-field theory, the coupling components  defining the effective singlet pairing interaction $g_{1\perp} + g_{2\perp}$ entering in $\tilde{\chi}_{\rm dFFLO}$ from (\ref{z_ssq}), are continuously altered by SDW correlations in the course of decreasing $\Lambda(\ell)$    (See also Fig.~\ref{gintra}). The initial values of couplings or antinesting modify this  energy scale dependent interference effect.  This indicates that in practice, the observation of the lack of universality in the anomalous upper critical field, as a function of the applied pressure in the Bechgaard salts for instance, would be a distinctive signature of fluctuation induced unconventional pairing in the material\cite{Fuseya12}.  On experimental side, there are some indications that this is indeed the case\cite{Lee02b}.

%=======================================================================================
\section{INTERCHAIN INTERACTIONS}

We now turn to the influence of interchain electron-electron repulsive interactions introducing a non zero  $\mathfrak{g}_{i,\alpha}$ in the interaction parameters (\ref{gialpha}) of the extended quasi-1D electron gas model\cite{Gorkov74,PeryleneSC1}.  We shall take for simplicity the transverse backward and forward scattering amplitudes equal by putting, $\mathfrak{g}_{1,\alpha}=  \mathfrak{g}_{2,\alpha}\equiv \mathfrak{g}$, for both parallel ($\alpha=\|)$ and perpendicular ($\alpha=\perp)$ spins. As for the transverse umklapp amplitude, we have the following  ratio with   backward scattering, $ \mathfrak{g}_{3\perp}/\mathfrak{g} = g_{3\perp}/g_{1\perp}$, which is the same as for   intrachain interactions discussed in Sec.~III~A. 

\begin{figure}  
\includegraphics[width=8cm]{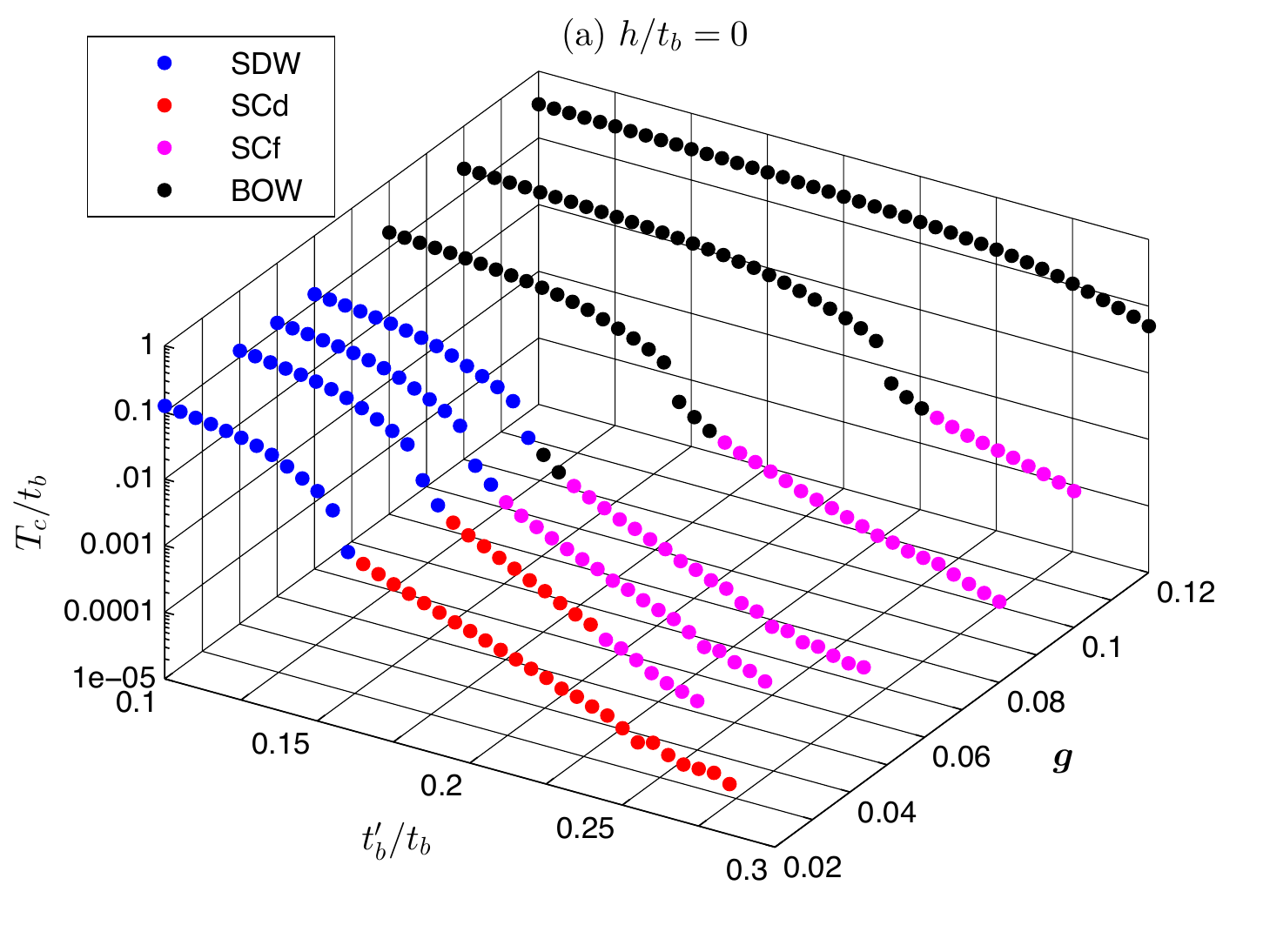}\\ \includegraphics[width=8cm]{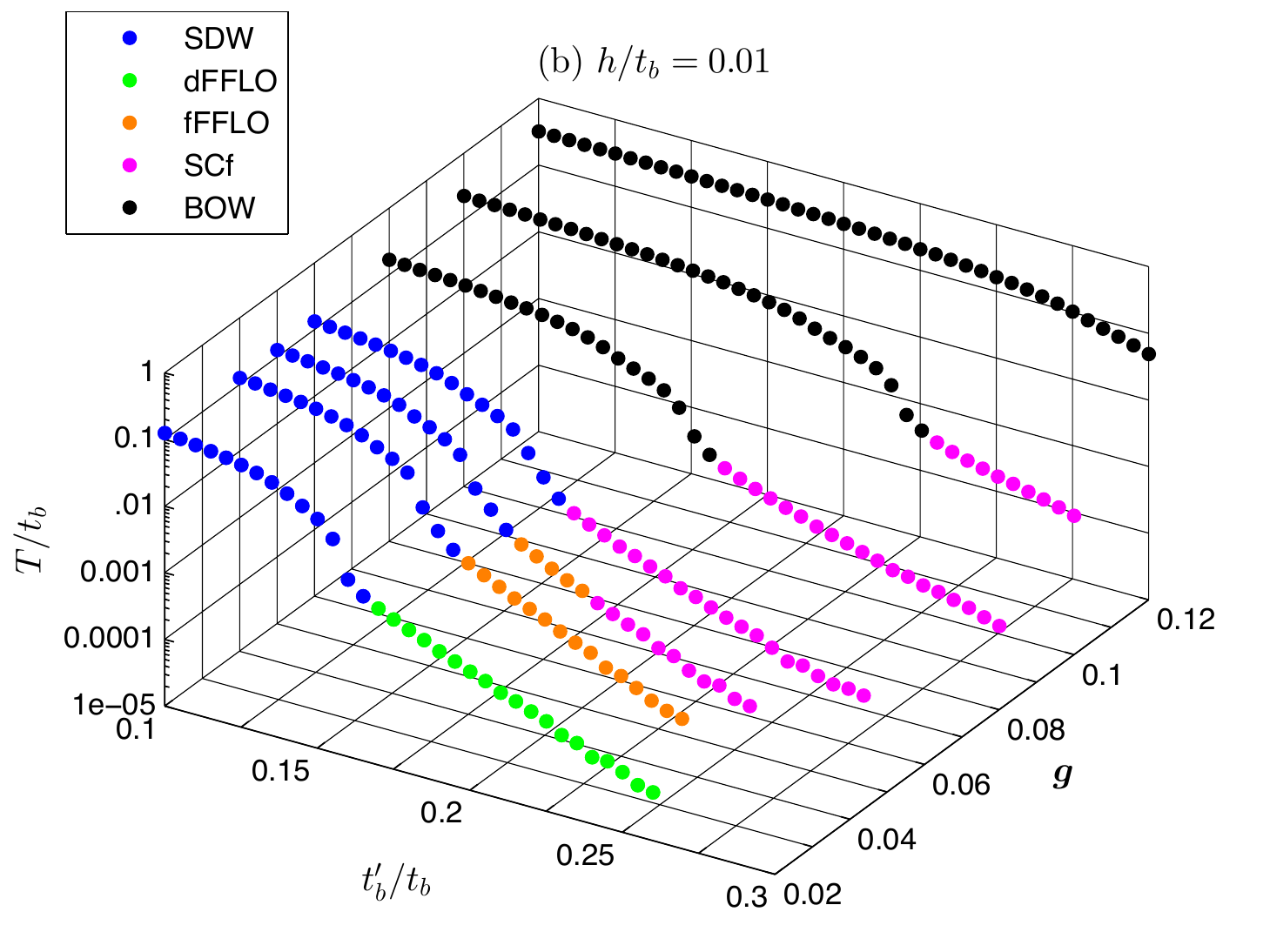}
\caption{ Phase diagram  of the extended quasi-electron gas model at (a):  $h=0$,  and (b): $h/t_b= 0.01$. }
\label{Phasesg}
\end{figure}

We first review the case in zero magnetic field, which was  examined by Nickel {\it et al.}\cite{PeryleneSC1}. By increasing the amplitude of $\mathfrak{g}$, the SDW$\to$SCd sequence of instabilities tuned by $t_b'$ is modified from the relatively small value, $\mathfrak{g}\simeq 0.04$, of interchain repulsion. According to Fig.~\ref{Phasesg}~(a), a triplet  $f$-wave  instability in $\tilde{\chi}_{\rm SCf}$ of  (\ref{SCtriplet0}-\ref{SCtriplet1}) gets into the sequence that becomes  SDW$\to$SCd$\to$SCf. By increasing  $\mathfrak{g}$, it transforms into SDW$\to$SCf, where  SDW  is connected directly to SCf at the quantum critical point $t_b'^*$. The emergence of SCf state emerges from the rise of BOW fluctuations which are detrimental to singlet SCd pairing. This sequence is soon modified by the incursion of a BOW instability from (\ref{CDWBOW}) in the sequence near $t_b'^*$, as shown in Fig.~\ref{Phasesg}~(a).  By increasing further  $\mathfrak{g}$, near 0.1, makes the SDW unstable and yields the sequence BOW$\to$SCf. This is also associated with the smearing of the quantum critical region to the benefit of the BOW ordering. Apart from few details at the quantitative level, the present results confirm those of Nickel {\it et al}\cite{PeryleneSC1}.

%~~~~~~~~~~~~~~~~~~~~~~~~~~~~~~~~~~~~~~~~~~~~~~~~~~~~~~~~~~~~~~~~~~~~~~~~~~~~~~~~~~~~~~~
%~~~~~~~~~~~~~~~~~~~~~~~~~~~~~~~~~~~~~~~~~~~~~~~~~~~~~~~~~~~~~~~~~~~~~~~~~~~~~~~~~~~~~~~

\begin{figure}
     \includegraphics[width=5.5cm]{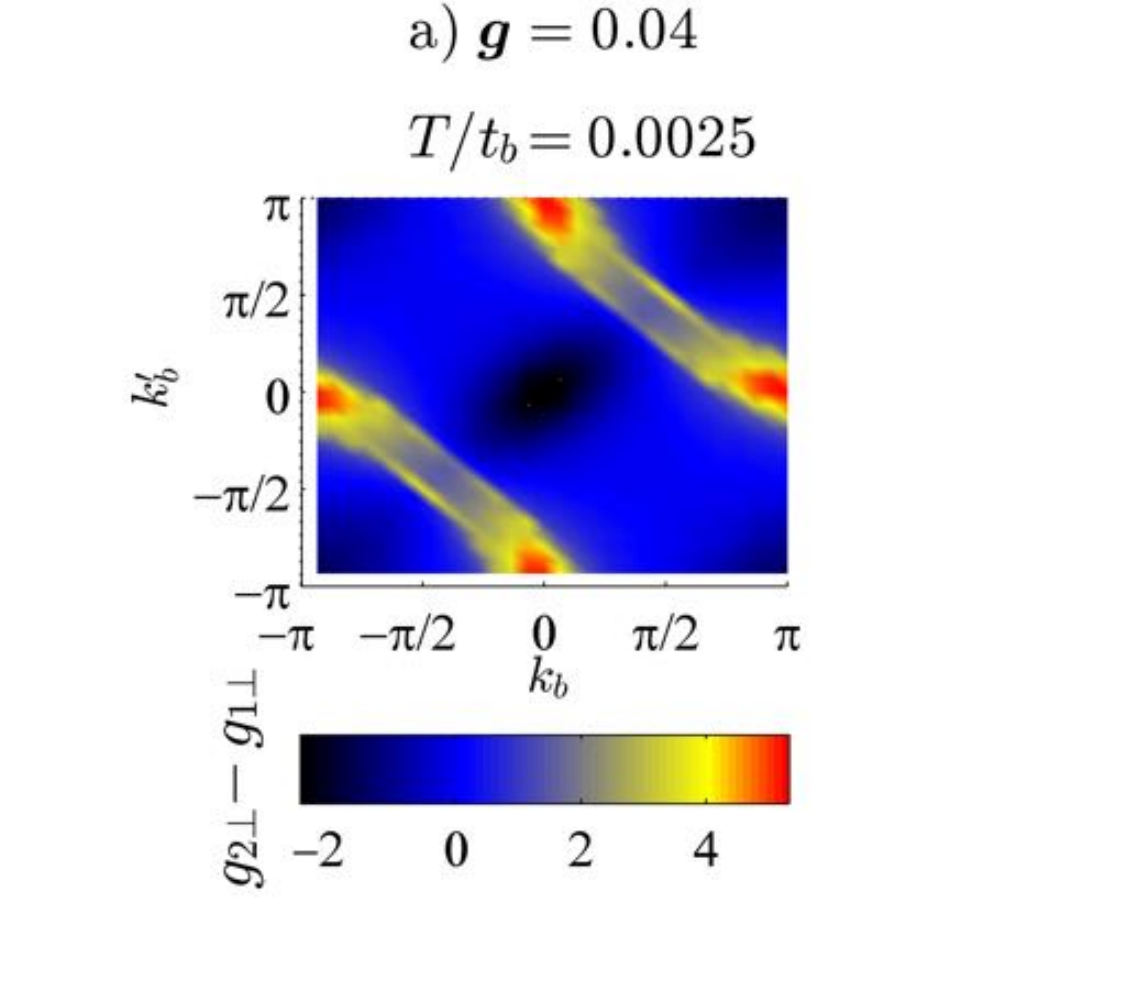} \\
     \includegraphics[width=5.5cm]{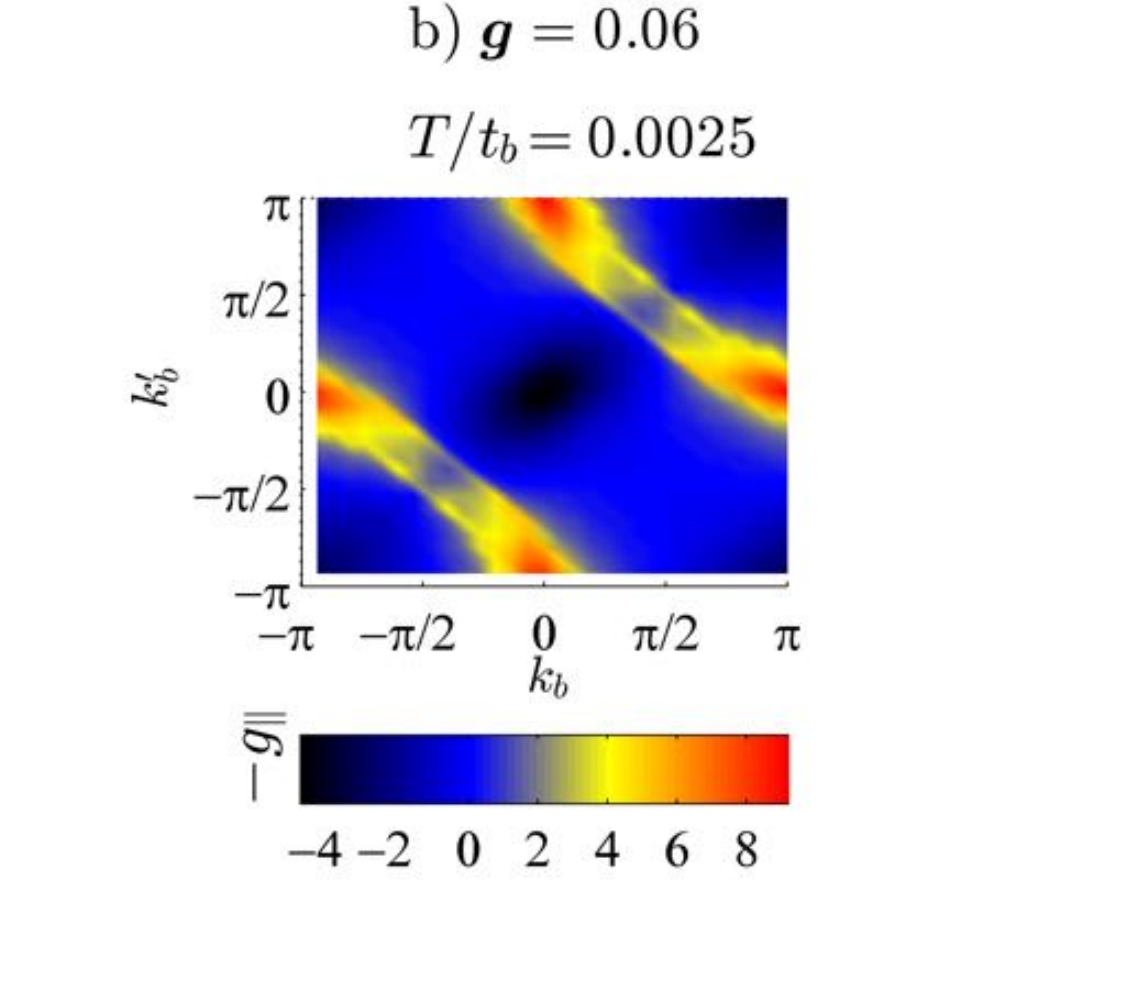}\\
     \includegraphics[width=5.5cm]{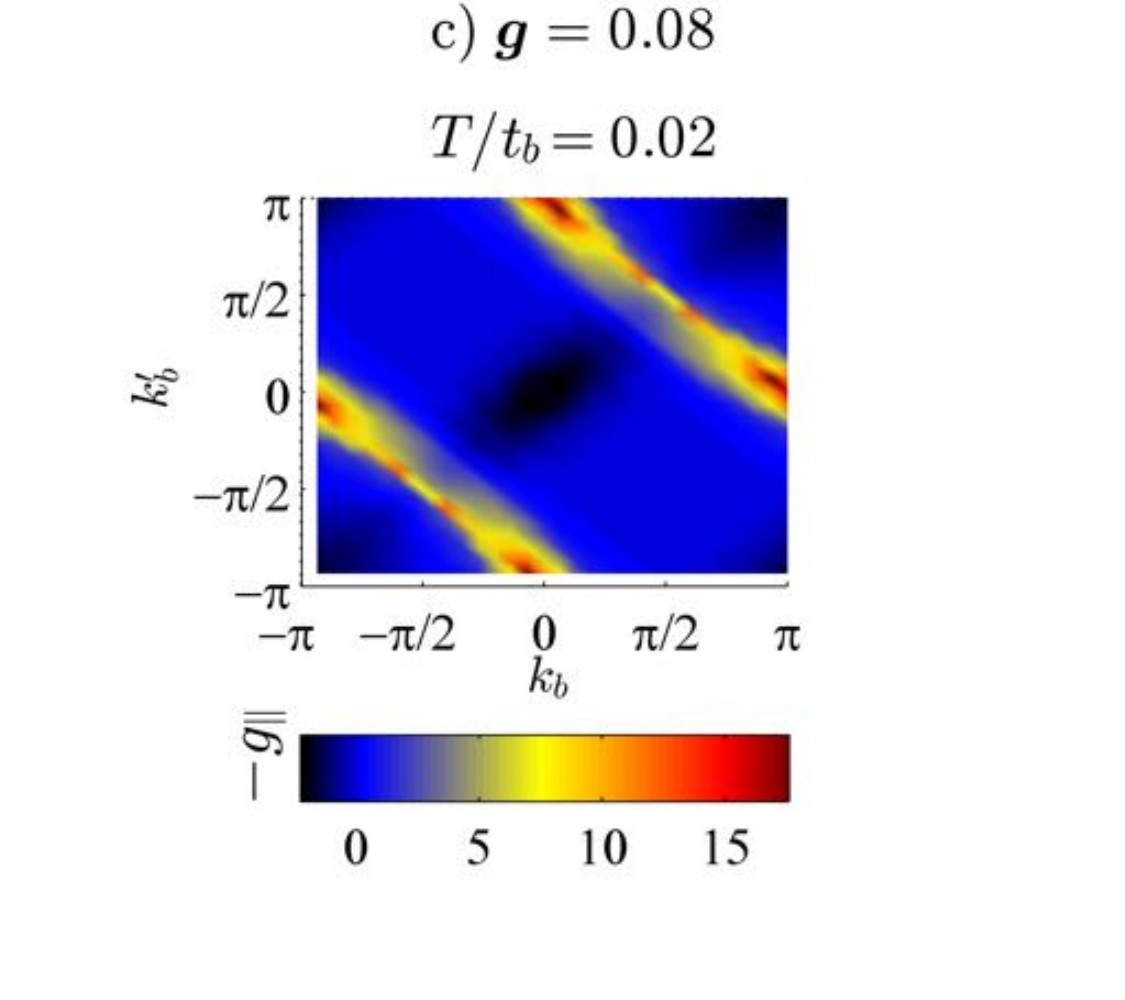} 
  \caption{Low temperature effective scattering amplitudes close to $T_c$ in the momentum space at $h/t_b=0.01$, for: (a) $g_{2\perp} (k_b',-k_b',k_b) - g_{1\perp} (k_b',-k_b',k_b)$  for fFFLO ; (b) and  (c) $g_{\|} (k_b',-k_b',k_b)$  for SCf$_1$. \label{contour_g}}
\end{figure}

If we now switch on the effect of magnetic field, we observe that the range of influence of triplet 
superconductivity is enlarged along the $\mathfrak{g}$ axis. Thus from $\mathfrak{g}\simeq 0.04$ and for $h/t_b \gtrsim 0.004$, the dFFLO state  of  Fig.~\ref{Phases_intra} becomes unstable against the formation of a  triplet fFFLO, $S_z=0$, state  governed by the divergence of (\ref{z_tsq}), as shown in Fig.~\ref{Suscg0}~(b) at $\mathfrak{g}=  0.04$. The related   combination of couplings $g_{2\perp}(k_b',-k_b',k_b) - g_{1\perp}(k_b',-k_b',k_b)$ in the $k'_bk_b$ plane is plotted in Fig.~\ref{contour_g}~(a) close to $T_c(h)$. One observes  pronounced modulations in momentum space compatible with $f_{\rm fFFLO}=\sqrt{2}\cos k_b^{(\prime)}$ and peaks at $k_b^{(')}=0,\pm \pi$, along the lines  $k_b'=- k_b \pm \pi$, which results mainly from  SDW  scattering. According to Fig.~\ref{Suscg0}~(b),  in this range of $\mathfrak{g}$,   SDW correlations are by far dominant down to very close to $T_c(h)$ and act as the main source of interchain pairing for the  fFFLO state at $S_z=0$.   It is worth mentioning that its  existence  has not been reported from  mean-field theory analysis \cite{Aizawa09}. However, the  FFLO mixing  with triplet superconductivity has been found from this analysis and from DMRG  in  the two-legs ladders systems at strong coupling \cite{Roux06}.
%~~~~~~~~~~~~~~~~~~~~~~~~~~~~~~~~~~~~~~~~~~~~~~~~~~~~~~~~~~~~~~~~~~~~~~~~~~~~~~~~~~~~~~~
%~~~~~~~~~~~~~~~~~~~~~~~~~~~~~~~~~~~~~~~~~~~~~~~~~~~~~~~~~~~~~~~~~~~~~~~~~~~~~~~~~~~~~~~

When $\mathfrak{g}$ further increases to reach $\mathfrak{g}\simeq  0.05$, the  fFFLO state becomes in turn unstable to the benefit of uniform triplet  SCf$_1$  state at $S_z=1$, and a sequence of instabilities, SDW$\to$SCf$_1$ along $t_b'$, as indicated in Fig.~\ref{Phasesg}~(b). This sequence in the above $\mathfrak{g}$ range is similar to the one found in Fig.~\ref{Phasesg}~(a) in the absence of field. Following (\ref{SCtriplet1}), the  SCf$_1$ pairing is directly connected to the combination  of couplings $g_\|(k_b',-k_b',k_b)$ for parallel spins. From  Fig.~\ref{contour_g}~(b), $g_\|$ presents strong modulations in the $k_b'k_b$ plane near $T_c(h)$, which are consistent with  the form factor $f_{\rm SCf}=\sqrt{2}\cos k_b^{(\prime)}$. As meant by the couplings involved in the BOW susceptibility in (\ref{CDWBOW}), peaks  along the lines  $k_b'=- k_b \pm \pi$ are consistent with the presence of strong BOW correlations  which acts as the main source of SCf$_1$ pairing\cite{PeryleneSC1,KurokiGr,Kajiwara09}. As displayed in Fig.~\ref{Suscg0}~(b), the amplitude of BOW correlations  are close in amplitude to SDW. 

At  $\mathfrak{g} \gtrsim 0.08$, the SDW state becomes in its turn unstable at low $t_b'$ in favour of a BOW state and the sequence   BOW$ \to$SCf$_1$ as a function of $t_b'$. The importance of BOW growth at the expense of SCf along the $\mathfrak{g}$ axis, as found in   the absence of field [Figs~\ref{Phasesg}~(a) and (b)]. This is reflected  in Fig.~\ref{contour_g}~(c) for the modulation of the relevant coupling, $g_\|$, for SCf$_1$  in the $k_b'k_b$ plane, which is less pronounced on the negative side.

%~~~~~~~~~~~~~~~~~~~~~~~~~~~~~~~~~~~~~~~~~~~~~~~~~~~~~~~~~~~~~~~~~~~~~~~~~~~~~~~~~~~~~~~

\begin{figure} 
 \includegraphics[width=8cm]{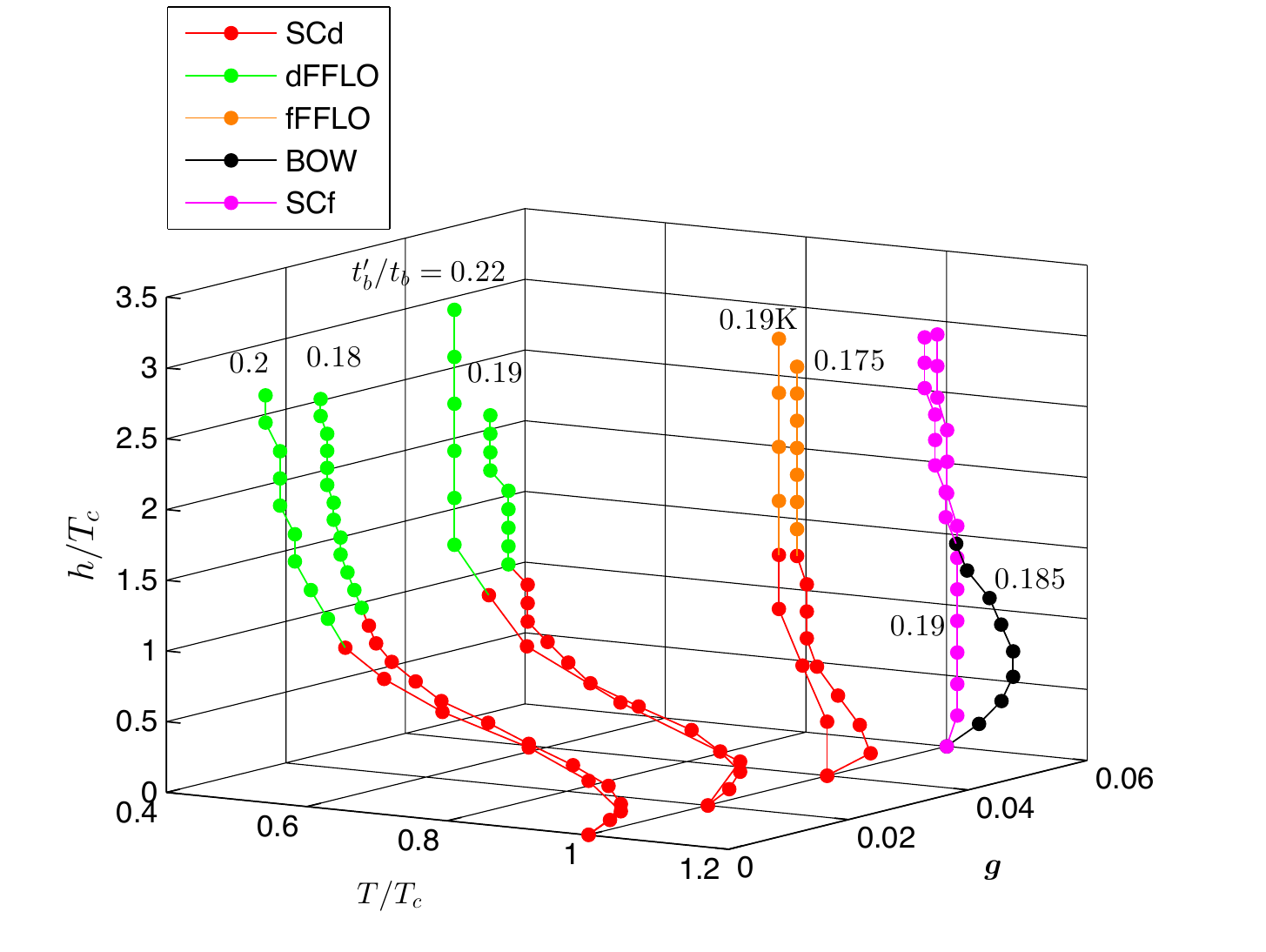}
 \caption{  The evolution of $h_{c2}$ $vs$ temperature  for various $t^\prime_b$ and  as a function of interchain coupling  $\mathfrak{g}$. \label{H-T} }
 \end{figure}

It is instructive to trace the temperature dependence of the critical field $h_{c2}(T)$ for the above selected ranges of interchain coupling $\mathfrak{g}$. At very low $\mathfrak{g}$,   the Fig.~\ref{H-T} shows that under field, we have  the expected sequence  of instabilities SCd$\to$dFFLO previously found  in Fig.~\ref{hc2_intra} at $\mathfrak{g}=0$. The violation of the Pauli limit in the dFFLO regime is dependent on antinesting and is reduced upon increasing $t_b'$. 
%~~~~~~~~~~~~~~~~~~~~~~~~~~~~~~~~~~~~~~~~~~~~~~~~~~~~~~~~~~~~~~~~~~~~~~~~~~~~~~~~~~~~~~~
%~~~~~~~~~~~~~~~~~~~~~~~~~~~~~~~~~~~~~~~~~~~~~~~~~~~~~~~~~~~~~~~~~~~~~~~~~~~~~~~~~~~~~~~
At higher $\mathfrak{g}$, when  fFFLO becomes possible, the crossover of  SCd to fFFLO under field is much more rapid and the violation of the Pauli limit, consequently more pronounced, with an almost vertical upturn of $h_{c2}$. Increasing further $\mathfrak{g}$,  the vertical rise of the $h_{c2}$ line for  SCf$_1$ does not lead to a crossover to another state, except for $t_b'$ close to the junction with BOW order, where one can start with a BOW state at low field and which is followed at sufficiently high field by a reentrant triplet SCf$_1$ state.

\begin{figure} 
 \includegraphics[width=9cm]{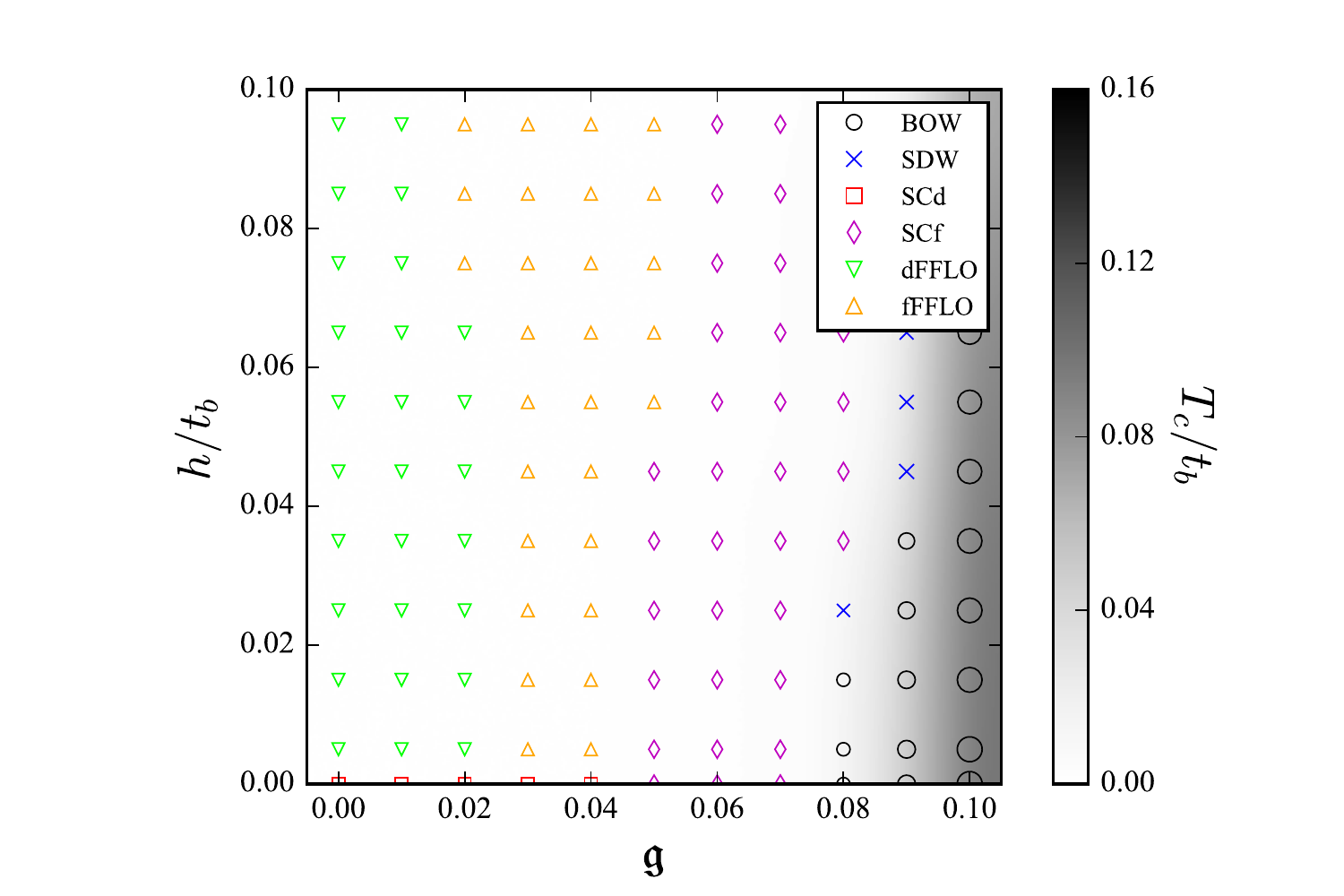}
 \caption{The phase diagram in the $\mathfrak{g}-h$ plane  for $t_b'/t_b = 0.2$. \label{H-g} }
 \end{figure}

By way of closing the section, we give in Fig.~\ref{H-g} the phase diagram  in the $\mathfrak{g}h$ plane which  displays  the transformation of ordered phases  under magnetic field  when the interchain interaction is varied at a fixed at $t_b'$ in the superconducting sector at ${\mathfrak{g}=0}$.  From the Figure, we observe that for a sizeable interval of weak repulsive $\mathfrak{g}$,  the possible modifications of superconductivity  expands under magnetic field   to the  benefit of  FFLO states. These are not exclusively restricted to the $d$-wave sector, but    also  belong to the triplet $f$-wave sector at $S_z=0$. 
 Thus at relatively weak interchain repulsion, the sequences SCd $\to$ dFFLO,   SCd $\to$ dFFLO $\to$ fFFLO and    SCd $\to$ fFFLO are possible transformations of superconductivity within an accessible range of magnetic field ($h/t_b < 0.1$). It is worth noting that no SCd $\to$ SCf  transition, from  singlet to triplet uniform superconductivity, is predicted  under field over  all the range of $\mathfrak{g}$ covered,
  this at variance with previous mean-field results\cite{Aizawa09}. Furthermore, as previously discussed for stronger interchain interaction, namely when instead of superconductivity, a BOW order is favoured   in zero field, the sequences BOW $\to$ SCf and BOW $\to$ SDW$_{xy}$ $\to$ SCf can be found showing the stabilization of uniform SCf from density-wave phases under sufficiently high magnetic field.

%~~~~~~~~~~~~~~~~~~~~~~~~~~~~~~~~~~~~~~~~~~~~~~~~~~~~~~~~~~~~~~~~~~~~~~~~~~~~~~~~~~~~~~~
%~~~~~~~~~~~~~~~~~~~~~~~~~~~~~~~~~~~~~~~~~~~~~~~~~~~~~~~~~~~~~~~~~~~~~~~~~~~~~~~~~~~~~~~
%=======================================================================================

\section{Summary and Concluding remarks}
In this work we employed the weak coupling renormalization group method to examine  the possible instabilities of the extended quasi-1D electron gas model with half-filling umklapp scattering and    the presence of    a magnetic field. The field is Zeeman coupled to spin degrees of freedom exclusively, which simulates the weakness of orbital pair breaking effect that characerizes specific planar field orientations  in  low dimensional superconductors  like the Bechgaard salts. 

For purely intrachain repulsive interactions, the SCd state of the electron gas is suppressed under field, evolving toward inhomogenous d-wave FFLO superconductivity rather than uniform triplet superconductivity. The dFFLO state is accompanied by a violation of the Pauli limiting field $H_P$ for singlet superconductivity, which is particularly enhanced by the constructive quantum interference between Cooper pairing and antiferromagnetic fluctuations. The enhancement is then found to scale  in a non universal way  with both the interaction and the distance to the quantum critical point joining superconductivity and the spin-density-wave state in the phase diagram. These results obtained  in the presence of half-filling umklapp scattering broaden the impact of  an earlier study made at incommensurate band filling\cite{Fuseya12}. 

For the extended version of the quasi-electron electron gas model when interchain Coulomb interaction is included and  a transition from d-wave to triplet $f$-wave superconductivity becomes possible. We find that its range of stability  is somewhat enlarged by the magnetic field as one moves along  the axis of interchain repulsion. The calculations reveal the existence of an intermediate $f$-wave FFLO state of zero total spin  projection, which emerges within a finite interval of interchain Coulomb repulsion  interaction before the onset of uniform $f$-wave superconductivity. 

The possible field-induced FFLO states obtained provide   an interesting avenue of interpretation for the persistant superconductivity in resistivity experiments observed  well above the Pauli limiting field of  the Bechgaard salts when they are very close to their quantum critical point. The results   also allow the possibility for a direct experimental test  of the theory  from future resistivity experiments that would be conducted on a large pressure interval, in order to check if the anomalous enhancement of the upper critical field is suppressed in the limit of high pressure, as predicted\cite{Fuseya12}. Existing resitivity  data on a (TMTSF)$_2$PF$_6$\cite{Lee02b}, though obtained in a limited range of pressures close to the critical value, head in  this direction.  

From the renormalization group viewpoint developed in the present work,  it is not clear a priori which one of the $d$ and $f$ FFLO states is likely to be more favorable  in systems like the Bechgaard salts. Both phases are occurring relatively  close one another and both are falling in a reasonable range of parameters for these materials.  It must be said, however, that the currently observed violation of the Pauli limit  in the Bechgaard salts  is significantly less pronounced than predicted in the  triplet case. This  would in turn tip the balance in favour of the singlet dFFLO scenario for the high field superconducting phase in the Bechgard salts.

\acknowledgments
We  thank Samuel Desrosiers for his help on computational aspects of this work. C.B. thanks the National Science and Engineering Research
Council of Canada (NSERC) under Grant No.~RGPIN-2016-06017, and the R\'eseau Qu\'eb\'ecois des Mat\'eriaux de Pointe
(RQMP) for financial support. Simulations were performed on
computers provided by Canadian Foundation for Innovation,
the Minist\`ere de l'\'Education des Loisirs et du Sport (Qu\'ebec),
Calcul Qu\'ebec, and Compute Canada.

\bibliography{/Users/cbourbon/Dossiers/articles/Bibliographie/articlesII.bib}

\end{document}